\documentclass[twocolumn, tighten]{aastex631}

\usepackage{graphicx,xcolor}	
\hypersetup{linkcolor=blue,citecolor=blue,filecolor=blue,urlcolor=blue}
\usepackage{amsmath}	
\usepackage{amssymb}	
\usepackage{newtxtext,newtxmath}

\shorttitle{Abundances in X-ray halos with RGS}
\shortauthors{Fukushima, Kobayashi, \& Matsushita}
\graphicspath{{./}}

\begin{document}

\title{O, Ne, Mg, and Fe Abundances in Hot X-ray-emitting Halos of Galaxy Clusters,
Groups, and Giant Early-type Galaxies with XMM-Newton RGS Spectroscopy}

\author[0000-0001-8055-7113]{Kotaro Fukushima}
\affiliation{Department of Physics, Tokyo University of Science, 1-3 Kagurazaka,
Shinjuku-ku, Tokyo 162-8601, Japan}
\correspondingauthor{Kotaro Fukushima}
\email{kxfukushima@gmail.com}

\author[0000-0001-7773-9266]{Shogo B. Kobayashi}
\affiliation{Department of Physics, Tokyo University of Science, 1-3 Kagurazaka,
Shinjuku-ku, Tokyo 162-8601, Japan}

\author[0000-0003-2907-0902]{Kyoko Matsushita}
\affiliation{Department of Physics, Tokyo University of Science, 1-3 Kagurazaka,
Shinjuku-ku, Tokyo 162-8601, Japan}




\begin{abstract}

Chemical elements in the hot medium permeating early-type galaxies, groups, and clusters
make them an excellent laboratory for studying metal enrichment and cycling processes
in the largest scales of the universe.
Here, we report the analysis by the XMM-Newton Reflection Grating Spectrometer of 14 early-type galaxies,
including the well-known brightest cluster galaxies of Perseus, for instance.
The spatial distribution of the O/Fe, Ne/Fe, and Mg/Fe ratios is generally flat
at the central 60\,arcsecond regions of each object, irrespective of whether
or not a central Fe abundance drop has been reported.
Common profiles between noble gas and normal metal suggest that the dust depletion process
does not work predominantly in these systems. Therefore, observed abundance drops
are possibly attributed to other origins, like systematics in the atomic codes.
Giant systems of high gas mass-to-luminosity ratio tend to hold a hot gas ($\sim$\,2\,keV) yielding
the solar N/Fe, O/Fe, Ne/Fe, Mg/Fe, and Ni/Fe ratios.
Contrarily, light systems at a subkiloelectronvolt temperature regime, including isolated or group-centered galaxies,
generally exhibit super-solar N/Fe, Ni/Fe, Ne/O, and Mg/O ratios.
We find that the latest supernova nucleosynthesis models fail to reproduce
such a super-solar abundance pattern.
Possible systematic uncertainties contributing to these high abundance ratios of cool objects
are also discussed in tandem with the crucial role of future X-ray missions.

\end{abstract}

\keywords{\href{http://astrothesaurus.org/uat/75}{Astrochemistry (75)};
\href{http://astrothesaurus.org/uat/224}{Chemical abundances (224)};
\href{http://astrothesaurus.org/uat/225}{Chemical enrichment (225)};
\href{http://astrothesaurus.org/uat/429}{Early-type galaxies (429)};
\href{http://astrothesaurus.org/uat/584}{Galaxy clusters (584)};
\href{http://astrothesaurus.org/uat/597}{Galaxy groups (597)};
\href{http://astrothesaurus.org/uat/813}{Intergalactic medium (813)};
\href{http://astrothesaurus.org/uat/847}{Interstellar medium (847)};
\href{http://astrothesaurus.org/uat/858}{Intracluster medium (858)}
}


\section{Introduction}
\label{sec:intro}

After primordial elements like H and He were produced by the Big Bang nucleosynthesis,
other heavy metals were synthesized in stars and ejected into interstellar space
by supernova (SN) explosions and mass-loss winds.
In general, core-collapse SNe (CCSNe) produce a significant fraction of O, Ne, and Mg \citep[e.g.,][]{Nomoto13}.
Type Ia SNe (SNeIa) dominantly forge the Fe-peak elements (Cr, Mn, Fe, and Ni)
in exploding white dwarves \citep[e.g.,][]{LN18}.
On the other hand, light metals like C and N are thought to be synthesized
in massive and/or asymptotic giant branch (AGB) stars
and dispersed via stellar mass loss \citep[e.g.,][]{Prantzos18, Kobayashi20}.
The bulk of these elements now reside within the interstellar medium
that pervades the intra- and intergalactic spaces.
In addition, the brightest cluster galaxies (BCGs) of clusters and groups, and early-type galaxies
would undergo an ongoing enrichment process by SNeIa and mass-loss winds.
Therefore, metal abundances and their spatial variation in the hot gaseous halos
\footnote{Throughout this paper, we use the term ``halo(s)'' interchangeably
referring to the intracluster medium, intra-group medium, and interstellar medium, for convenience.}
of these objects provide one of the most powerful probes
for investigating chemical enrichment at the largest scales of the universe.

In the last two decades, the advent of modern X-ray observatories offering both spectral
and spatial resolutions has allowed us to assess the cosmic mystery of metals from various aspects
\citep[e.g.,][for recent reviews]{Mernier18c, Gastaldello21}.
We can validate model calculations of the nucleosynthesis and/or mass-loss yield
by comparing them with the observed abundance pattern
\citep[e.g.,][]{dePlaa07, Hitomi17, Mao19, Simionescu19}.
The enrichment history, including metal transportation triggered by the cooling flows and feedback
from central galaxies, is also studied based on estimates of metal distributions, especially in central regions
\citep[e.g.,][]{Sanders16, Mernier22}. In particular, the detailed abundance pattern and distribution of elements
are well investigated for core regions of clusters and groups due to their high X-ray luminosities.

One intriguing conundrum is the ``metal abundance drop'' that is reported
within a central few-kiloparsec region of relaxed clusters and groups,
even though there should be active enrichment from their BCGs.
After the Fe abundance drop was first discovered in A2199 \citep{Johnstone02},
central abundance depletions have been reported in some cool-core systems
despite active element supply from their central galaxies
\citep[e.g.,][]{Churazov03, Panagoulia15, Mernier17, Liu19a}.
Based on a metallicity profile derived from CCD data,
\citet{Panagoulia15} attributed these drops partly to metal depletion into the cool dust grains,
which is supported by \citet{Lakhchaura19} for the Centaurus cluster since the noble gas Ar
shows a more marginal drop than the Si, S, and Fe abundances.
However, the abundance distribution of the same noble gas Ne has been poorly constrained
because CCD instruments installed on current observatories like Chandra and XMM-Newton
cannot sufficiently resolve the Ne~K-shell emission from the Fe-L bump \citep[][]{Liu19a}.

The current solution is the Reflection Grating Spectrometer (RGS) on board XMM-Newton
offering a much better resolution than CCDs.
While the RGS instrument has been initially aimed to observe point-like sources \citep{denHerder01},
it can provide important spatial information for extended emission from a luminous X-ray source
like a galaxy center \citep[e.g.,][]{Chen18, Zhang19}.
In addition, RGS surpasses any CCD detectors to date at resolving the spectra in the 0.8--1.5\,keV band
within which two major atomic codes, i.e., the atomic database (AtomDB, \citealt{Smith01}; \citealt{Foster12})
and the \textsc{spex} Atomic Code and Tables (SPEXACT, \citealt{Kaastra96}),
have been not fully convergent yet \citep[e.g.,][]{Gastaldello21}.
In this manner, \citet{Fukushima22} recently estimated the radial profile of the Ne/Fe ratio
in the X-ray-luminous Centaurus cluster using data with RGS.
They provided a robust result that the Ne/Fe ratio shows a flat distribution at the central part of the cluster
within which the Fe abundance drop has been reported.
On their flat RGS Ne/Fe and CCD Ar/Fe profiles, the abundance drop in the Centaurus core is
difficult to explain by the metal depletion process.
Similar flat profiles of Ne/Fe and Ar/Fe are also reported for M87 with CCD analysis \citep{Gatuzz23}.

Precise measurement of the O and Ne abundances and their distribution are also important
in order to screen the CCSN nucleosynthesis models more sophisticatedly
since these metals are forged by the C-ignition process
in supergiants \citep[e.g.,][]{Doherty15, Kobayashi20}.
While the O abundance is well investigated in clusters, groups, and early-type galaxies \citep[e.g.,][]{dePlaa17b},
few reports study and discuss detailed
(and robust) spatial variation and/or pattern of the Ne abundance due to the limitations of CCD detectors.
As mentioned above, however, RGS data can give us an excellent complement to our knowledge
of metal abundance with CCDs for luminous extended sources.

In this paper, we analyze the RGS data of the central part of galaxy clusters, groups, and early-type galaxies.
We mainly intend to study the N, O, Ne, Mg, Fe, and Ni abundance patterns and, if possible, their spatial distribution
in X-ray halos of these systems, which will provide useful information to assess abundance drops.
The uncertainties of the two latest atomic codes (AtomDB and SPEXACT)
are also discussed in our spectral analysis.
This paper is organized as follows. In Section\,\ref{sec:observation},
we summarize our XMM-Newton observations and data reduction.
In Section\,\ref{sec:results}, details of our spectral analysis are presented.
We interpret and discuss the results in Section\,\ref{sec:discussion}, and
present our conclusion summary in Section\,\ref{sec:conclusions}.
In this paper, cosmological parameters are assumed as $H_0=70$\,km\,s$^{-1}$\,Mpc$^{-1}$,
$\Omega_m=0.3$ and $\Omega_{\Lambda}=0.7$.
All abundances throughout this paper are relative to the proto-solar values of \citet{Lodders09}.
The errors are at 1$\sigma$ confidence level unless otherwise stated.

\section{Observations and Data reduction}
\label{sec:observation}

\begin{deluxetable*}{llrrrrr}
\tablenum{1}
\tablecaption{The RGS observations analyzed in this paper. \label{tab:observation}}
\tablewidth{0pt}
\tablehead{
\colhead{Object} & \colhead{ObsID} & \colhead{$l$, $b$} & \colhead{Redshift} &
\colhead{$N_\textrm{H}$} & \colhead{Luminosity} & \colhead{Total cleaned exposure} \\
\colhead{} & \colhead{} & \colhead{(deg)} & \colhead{} & \colhead{($10^{20}$\,cm$^{-2}$)}
& \colhead{($10^{10}$\,L$_\odot$)} & \colhead{(ks)}}

\decimalcolnumbers
\startdata
NGC~1404 & 0781350101 & $237.0, -53.6$ & 0.0065 & 1.6 & 5.0& 122.7  \\
NGC~4636 & 0111190701 & $297.7, 65.5$ & 0.0037 & 2.1 & 3.6 & 56.8 \\
NGC~4649 & 0021540201, 0502160101 & $295.9, 74.3$ & 0.0037 & 2.2 & 4.5 & 118.2 \\
NGC~5846 & 0021540101/501, 0723800101/201 & $0.42, 48.8$ & 0.0061 & 5.1 & 3.7 & 188.8 \\
M49 & 0200130101 & $286.9, 70.2$ & 0.0044 & 1.6 & 9.6 & 76.3 \\
HCG~62 & 0504780501/601, 0112270701 & $303.6, 53.7$ & 0.0140 & 3.8 & 2.1 & 134.2 \\
Fornax & 0400620101, 0012830101 & $236.7, -53.6$ & 0.0046 & 1.6 & 3.9 & 143.8 \\
NGC~1550 & 0152150101, 0723800401/501 & $191.0, -31.9$ & 0.0123 & 16 & 2.2 & 165.5 \\
M87 & 0803670601, 0200920101, 0114120101 & $283.8, 74.5$ & 0.0042 & 2.1 & 6.6 & 182.0 \\
A3581 & 0205990101, 0504780301/401 & $323.1, 32.9$ & 0.0214 & 5.3 & 4.0 & 126.7 \\
A262 & 0109980101, 0504780201 & $136.6, -25.1$ & 0.0161 & 7.2 & 2.1 & 57.3 \\
AS~1101 & 0147800101, 0123900101 & $348.3, -64.8$ & 0.0580 & 1.2 & 11 & 99.3 \\
AWM~7 & 0605540101, 0135950301 & $146.3, -15.6$ & 0.0172 & 12 & 2.6 & 150.4 \\
Perseus & 0085110101/201, 0305780101 & $150.6, -13.3$ & 0.0183 & 21 & 8.0 & 167.6 \\
\enddata
\tablecomments{Column (1): Object name. Column (2): Observation ID of XMM-Newton.
Column (3): Galactic coordinates of object centers.
Column (4): Redshift of each object retrieved from \citet{Chen07} and \citet{Snowden08}.
Column (5): Galactic absorption column densities calculated using the method of \citet{Willingale13}.
Column (6): The $B$-band luminosity of early-type galaxies and BCGs in clusters or groups
retrieved from \citet{Makarov14}.
Column (7): Total exposure time after the filtering procedures described in \S\,\ref{sec:observation}.}
\end{deluxetable*}

In this work, we analyzed the publicly archival observation data of core regions
of 14 nearby early-type galaxies, including BCGs in groups and clusters
with RGS mounted on XMM-Newton \citep{denHerder01}.
Table\,\ref{tab:observation} lists the analyzed objects in this work and some properties thereof.
Our sample is based on well-known objects of the chemical enrichment RGS sample
(CHEERS, \citealt{dePlaa17b}) for objects with long total exposure ($> 50$\,ks)
and good statistic of Fe-L lines ($> 10^5$\,counts in the 0.9--1.2\,keV band).
For simplicity, we limited the sample to ellipticals and/or BCGs.
Finally, there are four central galaxies of compact groups and seven BCGs in our sample,
including both types of systems with and without Fe abundance drops \citep{Panagoulia15, Liu19a}.
With this sample, we are able to study the Ne abundance of objects in various temperature and gas mass regimes.
The data reduction was performed using the XMM-Newton Science Analysis System (\textsc{SAS}).
All RGS data were reprocessed with \texttt{rgsproc} following the standard procedures
by the \textsc{SAS} team.
First, we extracted light curves and fitted them with Gaussians; then, the mean count rate $\mu$
and standard deviation $\sigma$ were calculated.
A threshold of $\mu \pm 2\sigma$ was applied to create good time intervals.
Total exposures after removing flare events are also summarized in Table\,\ref{tab:observation}.

When the X-ray peak of the target object locates near the pointing position,
the RGS spectra can offer one-dimensional spatial information over 5\,arcmin width
($\lesssim$ the width of the central CCD chip of the MOS detectors)
along the cross-dispersion direction \citep[e.g.,][]{Chen18, Zhang19, Fukushima22}.
All the data sets listed in Table\,\ref{tab:observation} have less than 0.5\,arcmin
offsets from the center of the target object.
In accordance with the prescription in \citet{Fukushima22}, we extracted first- and second-order RGS spectra 
centered on the emission peaks of each object along the dispersion direction.
The response matrices are generated for each spectrum.
We filtered events by the cross-dispersion direction and extracted
spectra over a wide region in each side with 0--6, 6--18, 18--60, 60--120, and 0--60\,arcsec.
As discussed in \citet{Fukushima22}, this method has the advantage
of region selection with an accurate width
compared to the use of the \texttt{xpsfincl} parameter that is traditionally performed.
The first- and second-order spectra are jointly fitted,
where the RGS1 and RGS2 spectra of each order are combined.
Whether or not we combined the different responses of these two instruments,
our analysis in the following sections provided consistent results.

\section{Analysis and Results}
\label{sec:results}

\subsection{Spectral fitting}
\label{subsec:fitting}

\begin{figure*}[htb!]
\plotone{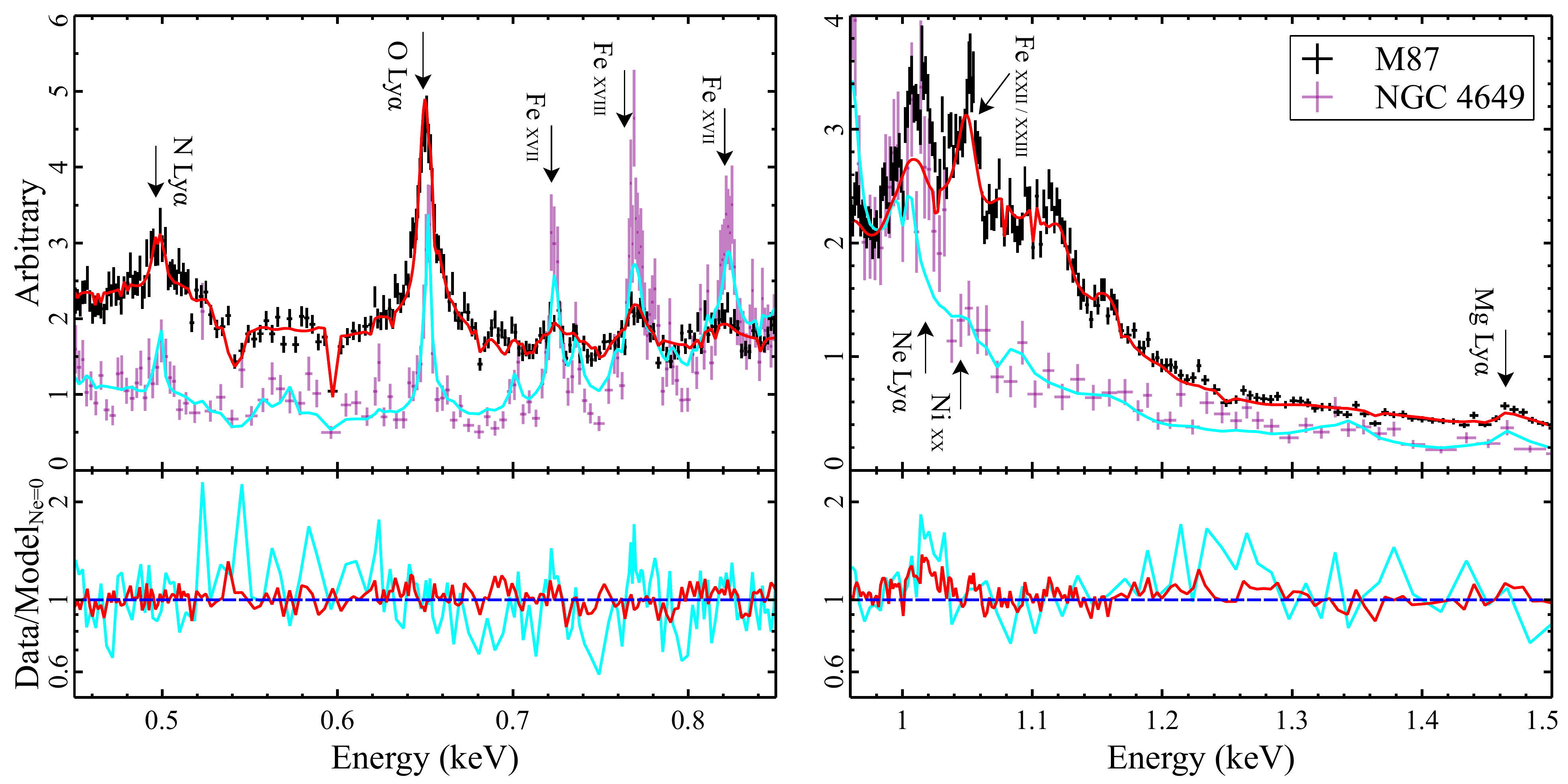}
\caption{RGS spectra of M87 and NGC~4649 extracted from the 0--60\,arcsec slice.
The crosses indicate the first-order data, and the solid lines represent the best-fit models
with AtomDB, but assuming Ne is 0\,solar.
The ratios of data to the zero-Ne models are plotted in the lower panels.
The emission lines of our interesting elements are marked on the upper panels.
\label{fig:spec}}
\end{figure*}

\begin{deluxetable*}{lrrrrrrc}
\tablenum{2}
\tablecaption{Best-fitting spectral parameters for halo emission within 60\,arcsec regions.
\label{tab:parameter}}
\tablewidth{0pt}
\tablehead{
\colhead{Object} & \colhead{$\langle kT \rangle$} & \colhead{$kT_\textrm{hot}$} &
\colhead{VEM$_\textrm{hot}$} & \colhead{VEM$_\textrm{middle}$} &
\colhead{VEM$_\textrm{cool}$} & \colhead{C-stat/dof} & \colhead{Regime} \\
\colhead{} & \colhead{(keV)} & \colhead{(keV)} & \colhead{($10^{10}$\,cm$^{-5}$)} &
\colhead{($10^{10}$\,cm$^{-5}$)} & \colhead{($10^{10}$\,cm$^{-5}$)} & \colhead{} & \colhead{}}

\decimalcolnumbers
\startdata
NGC~1404 & $0.49\pm 0.08$ & $0.68\pm 0.14$ & $16\pm 2$ & $16.9\pm 1.0$ & $1.7\pm 1.0$ & 4441.3/3512 & C \\
NGC~4636 & $0.61\pm 0.04$ & $0.77\pm 0.06$ & $27.6\pm 0.7$ & $14.4\pm 0.8$ & $2.4\pm 0.7$ & 3594.8/3259 & C \\
NGC~4649 & $0.82\pm 0.04$ & $0.86\pm 0.13$ & $27.2^{+2.7}_{-0.7}$ & $< 0.7$ & $1.8\pm 0.3$ & 3558.5/3343 & C \\
NGC~5846 & $0.56\pm 0.12$ & $0.8\pm 0.2$ & $20.1^{+0.5}_{-4.6}$ & $13.8^{+4.3}_{-0.6}$ & $4.0^{+0.4}_{-3.0}$ & 3468.9/3316 & C \\
M49 & $0.92\pm 0.05$ & $0.98\pm 0.08$ & $32.3\pm 0.9$ & $3.8\pm 0.3$ & $0.6\pm 0.5$ & 3738.9/3476 & C \\
HCG~62 & $0.80\pm 0.14$ & $1.0\pm 0.3$ & $26\pm 4$ & $5.7\pm 1.3$ & $4.3^{+9.7}_{-0.7}$ & 3240.8/3154 & C \\
Fornax & $0.83^{+0.17}_{-0.04}$ & $0.9^{+1.38}_{-0.06}$ & $50^{+28}_{-1}$ & $3.8\pm 0.4$ & $3.2^{+47.1}_{-0.7}$ & 4196.4/3582 & C \\
NGC~1550 & $1.26\pm 0.16$ & $1.3\pm 1.0$ & $40^{+3}_{-22}$ & $0.44^{+1.4}_{-0.2}$ & $0.6\pm 0.4$ & 3596.5/3575 & W \\
M87 & $1.5\pm 0.2$ & $1.53^{+1.1}_{-0.3}$ & $4.2^{+2.2}_{-0.4}\times 10^2$ & $16.6^{+8.7}_{-1.5}$ & $1.2^{+1.0}_{-0.5}$ & 4122.0/3672 & W \\
A3581 & $1.56\pm 0.07$ & $1.6\pm 0.2$ & $95\pm 5$ & $2.2\pm 0.8$ & $0.7\pm 0.4$ & 3649.5/3589 & W \\
A262 & $1.8\pm 0.3$ & $1.9\pm 0.4$ & $71\pm 10$ & $3.5\pm 0.5$ & $1.1\pm 0.3$ & 3181.4/3379 & W \\
AS~1101 & $2.6\pm 0.2$ & $2.6\pm 0.5$ & $(1.51\pm 0.14)\times 10^2$ & $3.1^{+3.2}_{-1.3}$ & $1.2\pm 0.3$ & 3812.1/3659 & H \\
AWM~7 & $2.8\pm 0.5$ & $2.9\pm 0.9$ & $79\pm 15$ & $1.3^{+4.8}_{-0.9}$ & $1.5\pm 0.3$ & 3672.6/3644 & H \\
Perseus & $2.73\pm 0.18$ & $2.8\pm 0.4$ & $(1.39\pm 0.13)\times 10^3$ & $11^{+45}_{-4}$ & $30\pm 4$ & 4359.0/3668 & H \\
\enddata
\tablecomments{Column (1): Object name. Column (2): The VEM-weighted average temperature.
Column (3): Temperature of the hot component.
Columns (4)(5)(6): The volume emission measures (VEMs) of each temperature component.
The VEM is given as $\int n_e n_\textrm{H} dV/(4\pi D^2)$,
where $V$ and $D$ are the volume of the emission region (cm$^3$)
and the angular diameter distance to the emitting source (cm), respectively.
Column (7): The ratio of C-statistic values to degrees of freedom.
Column (8): We define three temperature regimes of cool (C), warm (W), and hot (H):
$\langle kT \rangle$ is less than 1.0\,keV, greater than 1.0\,keV
and less than 2.0\,keV, and greater than 2.0\,keV, respectively.}
\end{deluxetable*}

\begin{deluxetable*}{lrrrrrr}
\tablenum{3}
\tablecaption{The Fe abundance and the N/Fe, O/Fe, Ne/Fe, Mg/Fe, and Ni/Fe ratios for halo emission
within 60\,arcsec regions. For NGC~4649, HCG~62, Fornax, and NGC~1550,
the O/Fe, Ne/Fe, and Mg/Fe ratios are computed by averaging over each radial bin
(see \S\,\ref{subsec:profile}). \label{tab:abund}}
\tablewidth{0pt}
\tablehead{
\colhead{Object}& \colhead{Fe} & \colhead{N/Fe} & \colhead{O/Fe} & \colhead{Ne/Fe} & \colhead{Mg/Fe} & \colhead{Ni/Fe} \\
\colhead{} & \colhead{(solar)} & \colhead{(solar)} & \colhead{(solar)} & \colhead{(solar)} & \colhead{(solar)} & \colhead{(solar)}}

\startdata
NGC~1404 & $0.34\pm 0.02$ & $2.1\pm 0.4$ & $0.58\pm 0.04$ & $1.28\pm 0.14$ & $1.66\pm 0.15$ & $3.4\pm 0.5$ \\
NGC~4636 & $0.56\pm 0.02$ & $1.6\pm 0.3$ & $0.64\pm 0.04$ & $0.89\pm 0.09$ & $1.10\pm 0.11$ & $2.4\pm 0.3$ \\
NGC~4649 & $0.39^{+0.08}_{-0.02}$ & $2.6\pm 0.8$ & $0.73\pm 0.09$ & $1.52\pm 0.12$ & $1.31\pm 0.14$ & $1.7\pm 0.4$ \\
NGC~5846 & $0.35\pm 0.01$ & $2.5\pm 0.5$ & $0.74\pm 0.09$ & $1.24\pm 0.13$ & $1.33\pm 0.16$ & $4.2\pm 0.6$ \\
M49 & $0.46\pm 0.03$ & $2.7\pm 0.6$ & $1.01\pm 0.09$ & $1.45\pm 0.19$ & $1.86\pm 0.16$ & $2.9\pm 0.4$ \\
HCG~62 & $0.24^{+0.10}_{-0.02}$ & $1.1\pm 0.8$ & $1.01\pm 0.07$ & $1.44\pm 0.18$ & $2.2\pm 0.2$ & $1.5\pm 0.7$ \\
Fornax & $0.28\pm 0.02$ & $4.3\pm 0.9$ & $1.23\pm 0.09$ & $1.52\pm 0.13$ & $1.77\pm 0.16$ & $4.8\pm 0.6$ \\
NGC~1550 & $0.57\pm 0.07$ & $3.5\pm 1.3$ & $1.31\pm 0.10$ & $0.91\pm 0.16$ & $0.61\pm 0.18$ & $1.9\pm 0.5$ \\
M87 & $0.67^{+0.16}_{-0.04}$ & $2.5\pm 0.4$ & $1.09\pm 0.05$ & $0.97\pm 0.07$ & $1.02\pm 0.08$ & $1.9\pm 0.2$ \\
A3581 & $0.66\pm 0.04$ & $< 1.3$ & $1.14\pm 0.10$ & $0.96\pm 0.17$ & $1.15\pm 0.15$ & $0.7\pm 0.3$ \\
A262 & $1.0\pm 0.2$ & $3.8\pm 1.4$ & $1.23\pm 0.16$ & $1.1\pm 0.3$ & $0.7\pm 0.3$ & $1.1\pm 0.6$ \\
AS~1101 & $0.67\pm 0.09$ & $< 0.8$ & $1.10\pm 0.10$ & $0.99\pm 0.14$ & $0.76\pm 0.15$ & $1.0\pm 0.4$ \\
AWM~7 & $2.0^{+1.4}_{-0.4}$ & $< 0.4$ & $1.17\pm 0.09$ & $0.8\pm 0.2$ & $1.33\pm 0.17$ & $1.3\pm 0.3$ \\
Perseus & $0.84\pm 0.06$ & $0.9\pm 0.5$ & $1.58\pm 0.05$ & $1.14\pm 0.07$ & $0.73\pm 0.05$ & $1.02\pm 0.13$ \\
\enddata
\end{deluxetable*}

The spectral fittings of the RGS X-ray spectra globally follow the same strategy of \citet{Fukushima22}.
We use \textsc{xspec} package version 12.10.1f \citep{Arnaud96}.
In the following spectral analysis, all spectra are fitted using the C-statistics \citep{Cash79}
to estimate the model parameters and their error range without bias \citep{Kaastra17}.
The spectra are re-binned to have a minimum of one count per spectral bin.
We jointly fit the first- and second-order spectra in the $0.45$--$1.75$\,keV and $0.8$--$0.75$\,keV bands,
respectively, with the same parameters.
Given that the halo emission is dominant in our analyzing regions
compared to the other astrophysical background emissions,
the possible instrumental component is taken into account by a steep power-law model.
We take the spectral broadening effect due to the spatial extent of the source
into account by using the \texttt{rgsxsrc} model with MOS1 images.

We use the latest AtomDB version 3.0.9 \footnote{\url{http://www.atomdb.org}}
to model emissions from collisional ionization equilibrium (CIE) plasmas.
The spectra of each object are modeled with triple \texttt{bvvtapec} components.
Two temperatures for line and continuum have the same value
because the continuum temperature is poorly constrained for the RGS spectra.
Although the double components model has been used to reproduce halo emissions
\citep[e.g.,][]{Panagoulia13,dePlaa17b, Lakhchaura19, Simionescu19},
it can still underestimate the Fe abundance in some systems (the ``double Fe bias'', see \citealt{Mernier22}).
Here, the triple components model is used in our analysis.
We do not adopt any assumption regarding the emission measures of each temperature component.
Instead, temperatures for each component are coupled along a geometric series
with a common ratio of 0.5 to reproduce the temperature structure
as $kT_\textrm{hot}:kT_\textrm{middle}:kT_\textrm{cool}=1:0.5:0.25$.
We note that an additional fourth component does not change fitting results
nor improve the fitting significantly, same as the Centaurus cluster in \citet{Fukushima22}.
The model with free temperatures also makes little change to other parameters.
Our modeling interpolates the continuous temperature distribution in halos sufficiently,
giving better fits than a Gaussian differential emission measure model.
All three CIE plasmas are modified by a common Galactic absorption of \texttt{phabs}
whose column densities are calculated following the method of \citet{Willingale13}
for each object (summarized in Table\,\ref{tab:observation}).
We adopt absorption cross sections of \citet{Verner96}.

Here, we show the representative spectra derived
from the 60\,arcsec cores of M87 and NGC~4649 (Figure\,\ref{fig:spec}).
Prominent K-shell emission lines of N, O, Ne, and Mg,
L-shell emission lines of Fe are clearly detected in the spectra.
In addition to these elements, the Ni abundance is also allowed to vary, and the abundances of other elements
whose emission lines are shrouded under the continuum emission are fixed to the solar value.
All abundances for three temperature components are tied together.
Finally, we allow the volume emission measures (VEMs) of each component to vary,
which is given as $\int n_e n_\textrm{H} dV/(4\pi D^2)$,
where $n_e$ and $n_\textrm{H}$ are electron and proton density, $V$ and $D$
are the volume of the emission region and the angular diameter distance to the object, respectively.

We perform the fitting to spectra extracted not only from each slice
with 0--6, 6--18, 18--60, and 60--120, and 0--60\,arcsec.
Our modeling largely gives a good fit with the ratios of the C-statistic value to degrees of freedom
(C-stat/dof)\,$\sim 1$ in the broader region as well as each slice.
The fitting results are summarized in Tables\,\ref{tab:parameter} and \ref{tab:abund}
for the 0--60\,arcsec regions. Results for each slice will be indicated in figures in subsequent sections.
In Figure\,\ref{fig:spec}, the model partially fails to reproduce the spectrum of NGC~4649
around the 0.7--0.8\,keV band, wherein the Fe~\textsc{xvii} and Fe~\textsc{xviii} lines are affected
by a large optical depth of resonant scattering \citep[e.g.,][]{Sanders08, Ogorzalek17}.
\citet{Mernier22} suggests that fearlessly excluding the Fe~\textsc{xvii} line
makes marginal changes in the abundance ratios less than 10 percent.
In addition, we also test to add Gaussians with a negative normalization at these emission lines
to avoid the resonant scattering effect in some early-type galaxies.
However, this model does not significantly change our measurement of abundance ratios.
It can matter just on reproducing line intensities of Fe,
especially for Fe~\textsc{xvii} and Fe~\textsc{xviii}, not on measuring abundances and ratios.

\subsection{Radial abundance profiles}
\label{subsec:profile}

\begin{figure*}[htb!]
\plotone{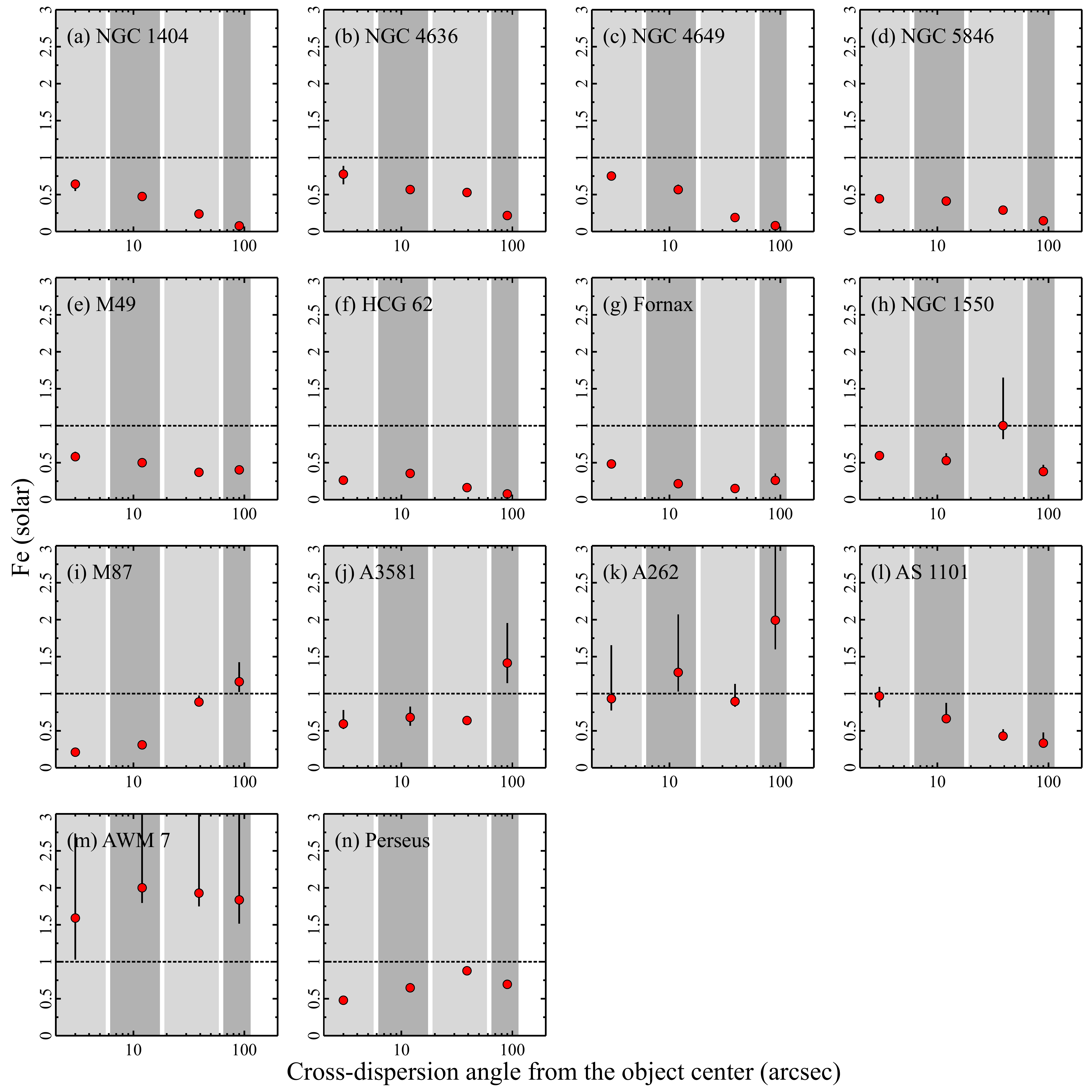}
\caption{Radial profiles of the Fe abundance as a function of radial distance in units of arcseconds
from the center of each object.
Grey areas show the width of the slice region for RGS analysis:
0--6, 6--18, 18--60, and 60--120\,arcsec from left to right, respectively.
The solar ratio is indicated with dotted lines on each panel.
\label{fig:feprofile}}
\end{figure*}

\begin{figure*}[htb!]
\plotone{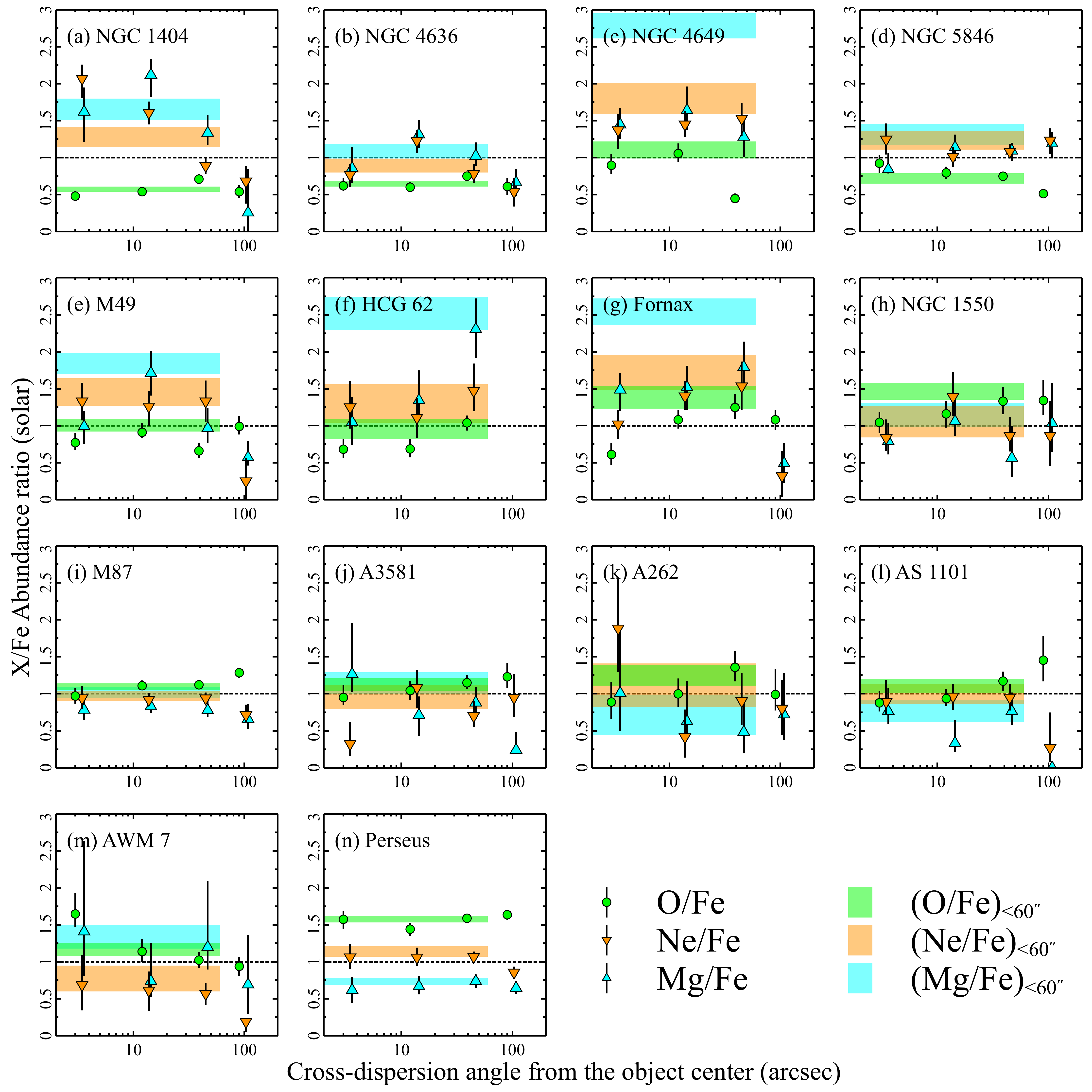}
\caption{Radial profiles of the O/Fe, Ne/Fe, Mg/Fe abundance ratios for each object.
The circles, down-, and up-triangles represent the O/Fe, Ne/Fe, and Mg/Fe ratios, respectively,
as a function of distance in units of arcseconds from the center of each object.
The plots are slightly shifted on the $x$-axis for clarity.
Color-shaded areas show the ratios obtained from broad 60\,arcsec slices.
The solar ratio is indicated with dotted lines on each panel.
\label{fig:abundprofile}}
\end{figure*}

Here, we show the central metal distribution in the hot halos of each system.
In Figure\,\ref{fig:feprofile}, radial profiles of the Fe abundance are plotted for each object.
In some previous works with CCD detectors,
NGC~4636, NGC~4649, NGC~5846, HCG~62, A3581, A262, AS~1101, and Perseus
are proposed to show an Fe abundance drop \citep[e.g.,][]{Churazov03, Panagoulia15, Liu19a}.
Our RGS results show no clear abundance declination in these objects,
excluding a hint of drop for the Perseus cluster.
M87 shows a sharp Fe drop, where no abundance drop
has been reported with CCD studies to date \citep[e.g.,][]{Matsushita03, Million11, Gatuzz23}.
However, we should be cautious in interpreting these profiles
because the metal abundances are estimated with the RGS instrument.
Additionally, the abundance drop might be detected only with broadband spectra,
including both the Fe~L- and K-shell emission lines \citep[][for the Centaurus cluster]{Gatuzz22b}.

The relative abundance ratio to Fe is a more robust parameter
than the absolute abundance in the RGS analysis \citep[e.g.,][]{dePlaa17b, Gastaldello21}.
We plot the O/Fe, Ne/Fe, and Mg/Fe abundance ratios in Figure\,\ref{fig:abundprofile},
discarding the outermost bins of NGC~4649 and HCG~62 with large errors ($>$\,1.5\,solar)
due to their line-poor or line-less spectra.
This is the first report of the Ne distribution using the RGS data sets.
Given that the Ne-Ly$\alpha$ lines are completely shrouded in the Fe-L bump with CCD spectra,
our Ne profile is possibly more reliable than previous ones \citep[e.g.,][]{Million11,Liu19a}.
More importantly, such a flat distribution has no individuality among the samples,
whether or not an object shows the abundance drop (in CCD spectra, e.g., \citealt{Panagoulia13, Liu19a}).
The ratios are largely around the solar ratio and give a flat profile
within the 60\,arcsec regions in each object but for NGC~1404.
In NGC~1404, the Ne/Fe and Mg/Fe ratios exhibit clear central enhancements
despite a flat distribution of O/Fe like the other systems.
Following \citet{Mernier22}, we also took the halo emission from the Fornax cluster
into account as a background in the outer two bins.
Our background estimation does not change the abundance measurements
in the outer region of NGC~1404.

\subsection{Abundances and temperature in the central 60\,arcsec}
\label{subsec:core}

As reported in \S\,\ref{subsec:profile}, the O/Fe, Ne/Fe, and Mg/Fe profiles
generally show a flat distribution in the central 60\,arcsec core of each halo,
suggesting that the metal content is remarkably uniform in the central cores.
The best-fit parameters for the 0--60\,arcsec slice are summarized
in Tables\,\ref{tab:parameter}, \ref{tab:abund}, and Figure\,\ref{fig:abundprofile} (shaded areas).
It is noteworthy that our estimated abundance ratios for the Perseus cluster are globally consistent
with the previous work by \citet{Simionescu19} despite a different fitting strategy.
The lower Fe abundance of theirs ($\sim 0.4$\,solar)
mainly originates from a narrower width of about 0.8\,arcmin than ours
and/or simple two-temperature components modeling.

In Table\,\ref{tab:parameter}, we also give the VEM-weighted average temperatures
(denoted as $\langle kT \rangle$) derived within 60\,arcsec.
These values suggest that our samples seem to be divided into three subgroups
by $\langle kT \rangle$ of halos.
Henceforward, two threshold temperatures ($\langle kT \rangle$\,$=$\,1 and 2\,keV)
determine three regimes of cool, warm, and hot halos (denoted by C, W, and H, respectively).
Tables\,\ref{tab:observation} and \ref{tab:parameter} suggest that three temperature regimes
correspond well to the traditional object classification, i.e., ``galaxies'', ``groups'', and ``clusters''.

One caveat is that the abundance ratios derived from the cumulative spectra are sometimes overestimated,
especially for Mg/Fe, compared to those of each individual bin
(e.g., see the cyan area and up-triangles for NGC~4649, Figure\,\ref{fig:abundprofile}(c)).
Such a schism is seen in the C-regime objects that have relatively low Mg lines (Figure\,\ref{fig:spec}(b)).
Abundance estimation by RGS is hard for elements
whose emission lines do not dominate the continuum \citep[e.g.,][]{dePlaa17b, Mao19, Gastaldello21}.
A strategy of narrow-band fits at $1.3$--$1.6$\,keV around the Mg~Ly$\alpha$ line,
which can minimize such biases, improves overestimation,
yielding Mg/Fe\,$=$\,1.5, 2.1, and 2.2\,solar for NGC~4649, HCG~62, and Fornax, respectively.
Thus, despite better photon statistics of cumulative spectra,
we report in Table\,\ref{tab:abund} and adopt in subsequent sections
the average ratios over the inner three bins instead of unnaturally high ratios for concerning objects.

\subsection{AtomDB vs SPEXACT}
\label{subsec:atomcode}

\begin{figure*}[htb!]
\plotone{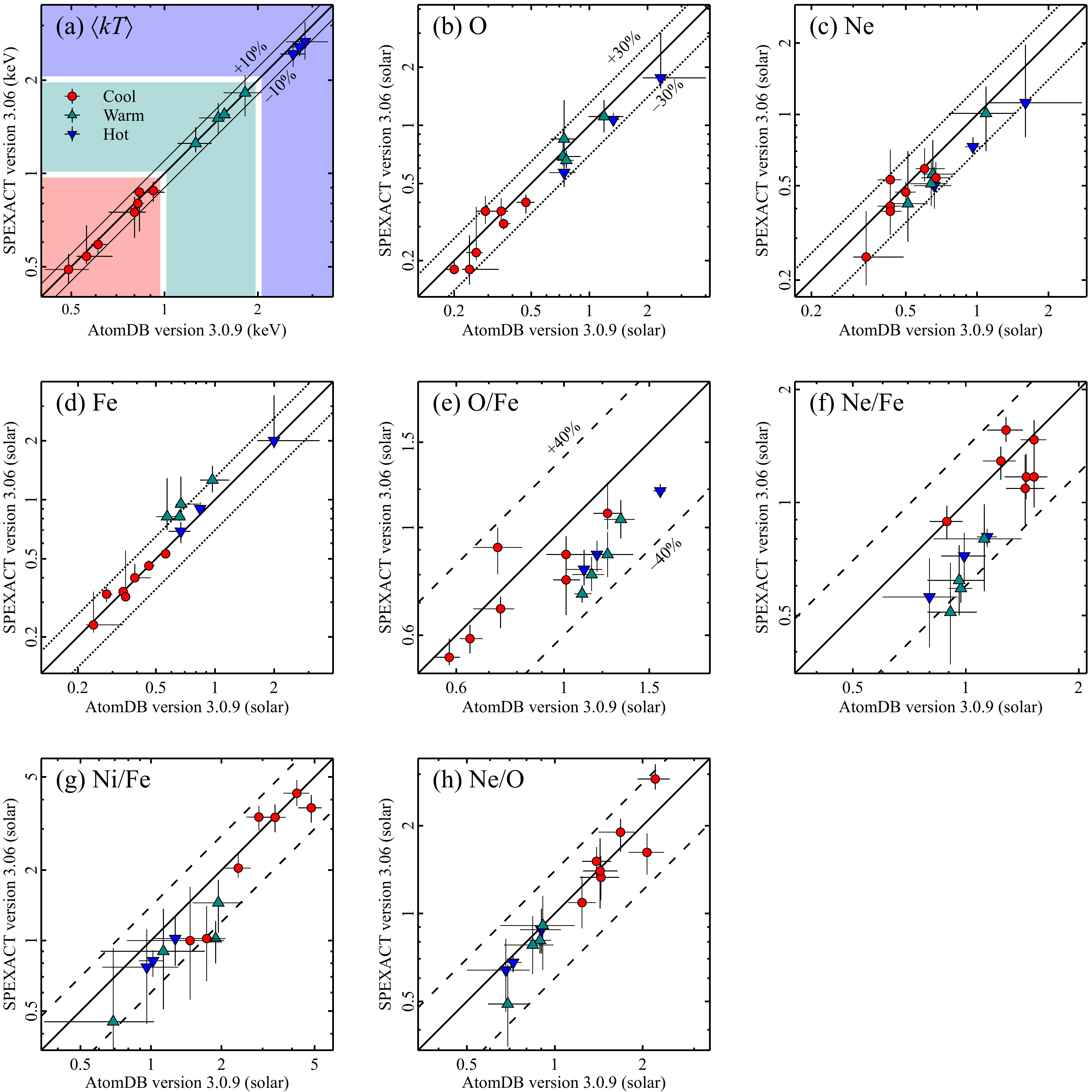}
\caption{Comparisons of the spectral parameters obtained from central 60\,arcsec slices
for each object when using AtomDB or SPEXACT.
(a) The VEM-weighted averaged halo temperature. Our sample is subdivided into
three subsamples: cool, warm, and hot (see the text) plotted by the circles, up-, and down-triangles, respectively.
(b, c, d) The absolute abundances of O, Ne, and Fe.
(e, f, g, h) The relative abundance ratios of O/Fe, Ne/Fe, Ni/Fe, and Ne/O.
The solid lines in each panel indicate equal values between the two codes.
The thin-solid, dotted, and dashed lines delimitate $\pm$10, 30, and 40 percent limits, respectively.
\label{fig:atomdata}}
\end{figure*}

Abundance measurements suffer from a systematic bias caused by atomic code uncertainties
\citep[e.g.,][for a review]{Mernier20a}.
In this work, we test the SPEXACT version 3.06 \citep{Kaastra96} for the spectra
extracted within 60\,arcsec regions to study the uncertainties
arising from systematics in the atomic data.
In order to minimize modest deviations in the modeling procedure of \textsc{xspec} and \textsc{spex},
we tabulate the lines and continuum calculated by a simple \texttt{cie} model
at 201 bins of temperature on \textsc{spex} and export them into APEC table format
that can be read in \textsc{xspec}.
This process is done with a \textsc{python} code
\footnote{\url{https://github.com/jeremysanders/spex_to_xspec}}.

In Figure\,\ref{fig:atomdata}, we show direct comparisons of $\langle kT \rangle$, abundances,
and relative abundance ratios between AtomDB and SPEXACT.
Clearly seen is that the $\langle kT \rangle$ measurements are in quite good agreements
between the two codes ($\lesssim$\,10 percent)
irrespective of halo temperature regimes (Figure\,\ref{fig:atomdata}(a)).
Such agreements on gas temperature between two codes have also been reported not only for RGS data
but also for calorimetric ones \citep{Hitomi18a, Simionescu19}.
The O, Ne, and Fe abundances are largely consistent with each other
in up to $\sim$\,30 percent differences.
For the abundance ratios, more moderate agreements $\lesssim$\,40 percent are
measured (Figures\,\ref{fig:atomdata}(e), (f), (g), and (h)).
The Ne/O ratios show a relatively small scatter compared to O/Fe, Ne/Fe, and Ni/Fe,
and are consistent within error ranges.
These differences between the two codes are possibly consistent with
the report of the abundance measurements in the intracluster medium \citep{Mernier20a}.
The most important is that these agreements between the atomic codes
are independent of the temperature regimes.

\section{Discussion}
\label{sec:discussion}

\subsection{Flat O/Fe, Ne/Fe, and Ne/O profiles}
\label{subsec:flat}

\begin{figure*}[htb!]
\plotone{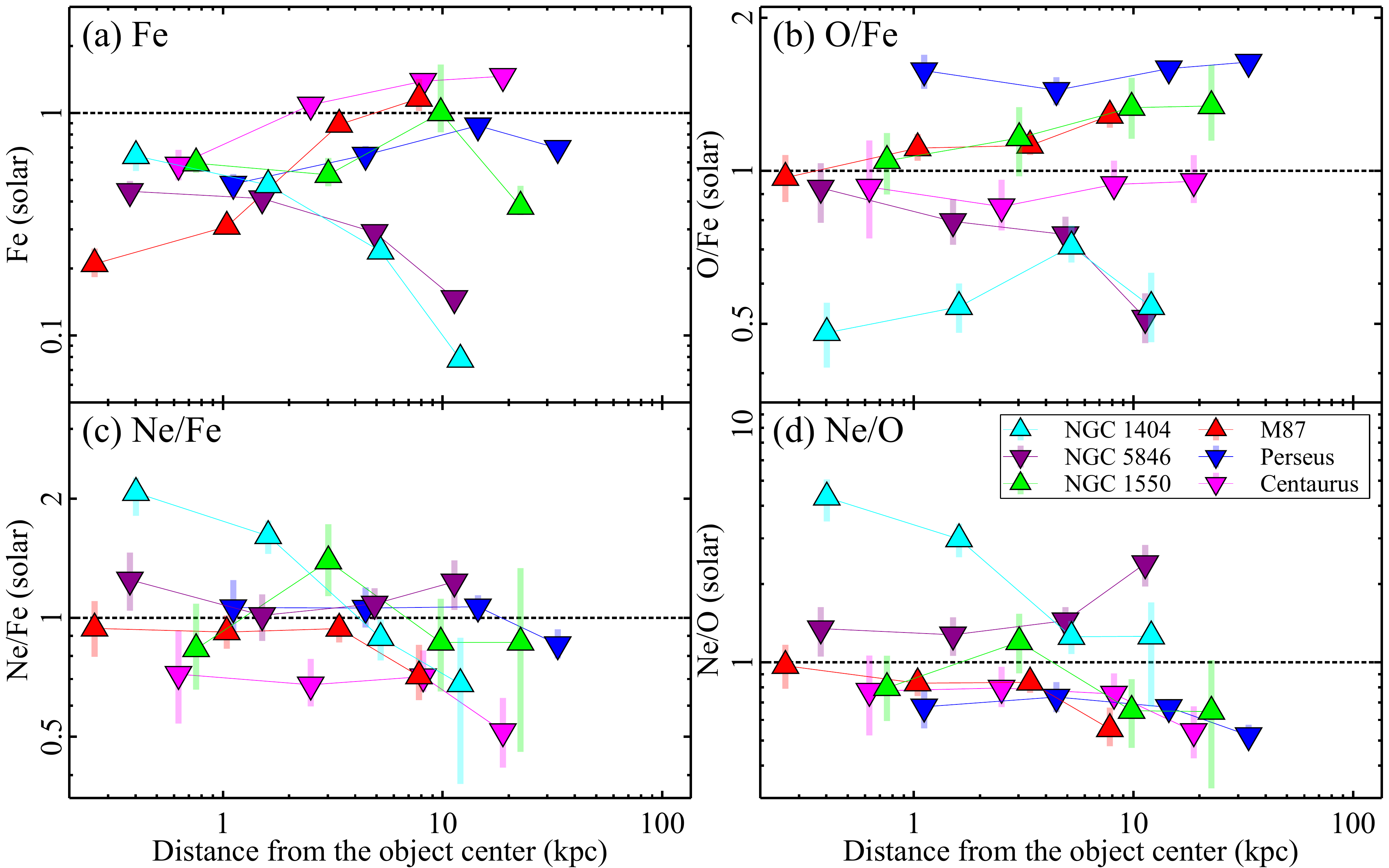}
\caption{Radial profiles of (a) Fe, (b) O/Fe, (c) Ne/Fe, and (d) Ne/O for representative objects.
All are plotted as a function of distance in units of kiloparsecs from the center of each system.
No Fe abundance drop has been reported in objects plotted by up-triangles,
and down-triangles are for the Fe drop ones.
The values of the Centaurus cluster are retrieved from \citet{Fukushima22}.
The solar ratio is indicated with dotted lines on each panel.
\label{fig:selectprofile}}
\end{figure*}

\begin{figure}[htb!]
\plotone{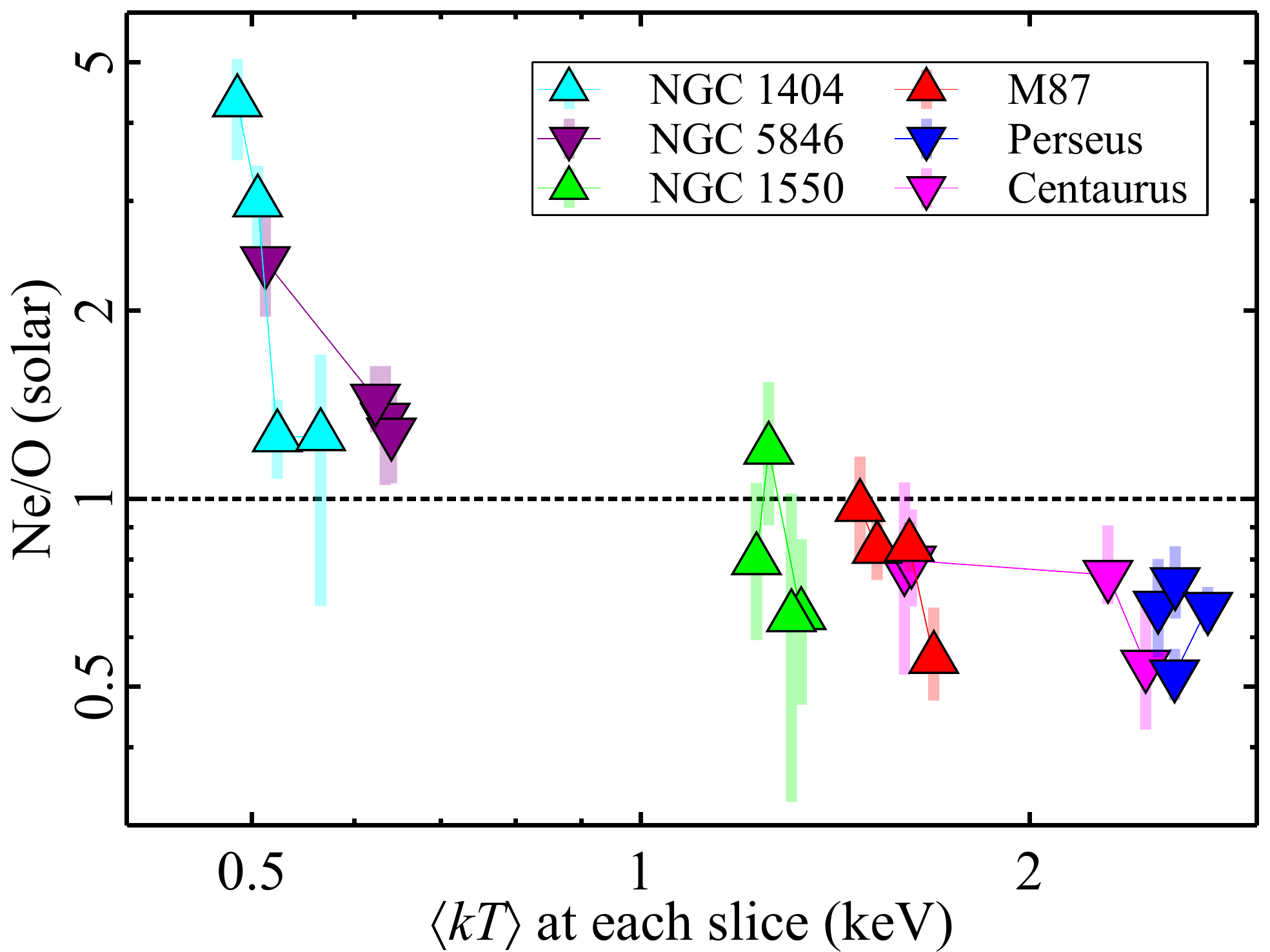}
\caption{The Ne/O ratio plotted against local $\langle kT \rangle$ at each slice
 for the same objects of Figure\,\ref{fig:selectprofile}.
The solar ratio is indicated with dotted lines on each panel.
\label{fig:neo}}
\end{figure}

As described in Section\,\ref{sec:intro}, the mechanisms of the abundance depletions are still elusive.
\citet{Panagoulia15} and \citet{Lakhchaura19} proposed the dust grains origin
for such abundance drop and noted that the distribution of non-reactive elements (Ne and Ar)
is important to inspect this picture.
For the Centaurus cluster, \citet{Fukushima22} reported flat radial profiles of the Ne/Fe and Ar/Fe ratios,
which suggests that Fe and non-reactive elements have the same spatial distribution.
This result is difficult to explain only by the dust origin process.

In Figures\,\ref{fig:selectprofile}, we show the (a) Fe, (b) O/Fe, (c) Ne/Fe,
and (d) Ne/O profiles for representative objects in addition to our previous results of the Centaurus cluster.
While the Fe abundance shows various profiles among the objects,
the O/Fe, Ne/Fe, and Ne/O abundance ratios are relatively flat in the central two or three bins,
except for NGC~1404.
For the majority of our samples, noble gas Ne and other metals identically distribute in halos
regardless of the (possible) presence of an abundance drop.
While O and Fe are independently synthesized by CCSNe and SNeIa, respectively,
the dominance of O and Ne must share the same origin.
Thus, flat Ne/Fe and Ne/O profiles strongly contradict the dust grains scenario.
Additionally, measuring the absolute abundance with the RGS spectra suffers
from systematic uncertainties since the grating spectra are dominated by line emissions
rather than continuum emissions \citep[e.g.,][]{Mernier18c}.
Hence, the Fe abundance drop suggested for M87 and Perseus
is caused at least partly by other physical processes
and/or uncertainties in abundance measurement like atomic codes.

One exception is a clear enhancement of Ne with respect to O and Fe for NGC~1404
(Figures\,\ref{fig:selectprofile}(c) and (d)).
However, \citet{Mernier22} determined the abundance distribution of Mg, Si, and Fe
by CCDs on board XMM-Newton, where the centrally-peaked profiles
rather than dropping ones are reported, especially for Fe.
Our increasing profiles of Fe and Mg/Fe toward the center support their results
(see Figures\,\ref{fig:feprofile}(a) and \ref{fig:abundprofile}(a)).
Additionally, if dust grains do work significantly in this object,
common increases of Ne/Fe and Mg/Fe are unreasonable
because Mg is easily depleted into dust compared to noble-gas Ne.
The dust formation scenario likely does not matter in NGC~1404 even if an increasing Ne/O profile is observed.

Worth mentioning is that the innermost $\langle kT \rangle$ of NGC~1404 is the coolest one
in our analysis (\S\ref{subsec:fitting}).
Here, we also plot the Ne/O ratios as a function of local $\langle kT \rangle$
at each slice (Figure\,\ref{fig:neo}).
Interestingly, high Ne/O ratios are observed at lower temperature slices in one object,
as well as for cooler systems among the sample.
The possible correlation of Ne/O to $\langle kT \rangle$ suggests some uncertainties
in the abundance estimation of cool plasmas.
This will be discussed in more detail in \S\,\ref{subsubsec:highne}.

\subsection{Metal composition, $\langle kT \rangle$ and $M_{<60''}/L_B$}
\label{subsec:comp}

\begin{figure*}[p!]
\plotone{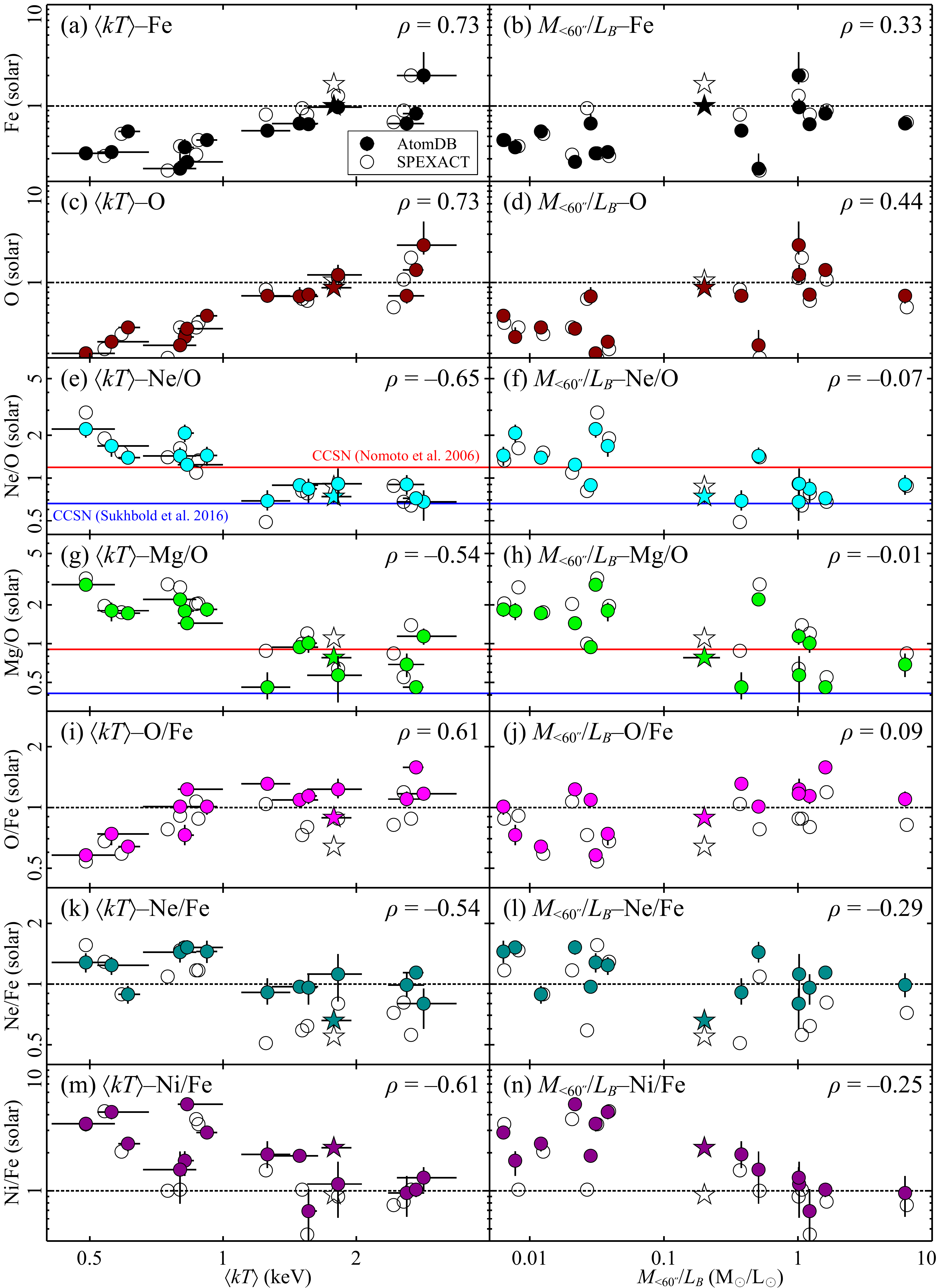}
\caption{The Fe and O abundances and the Ne/O, Mg/O, O/Fe, Ne/Fe, and Ni/Fe ratios as functions of
(a, c, e, g, i, k, m) $\langle kT \rangle$  and (b, d, f, h, j, l, n) $M_{<60''}/L_B$, respectively.
The filled and open circles represent the results with AtomDB and SPEXACT, respectively.
The stars indicate the results of Centaurus \citep{Fukushima22}.
Errors are shown only for the AtomDB results for viewing purposes.
We also show expected ratios by two major CCSN nucleosynthesis models
\citep{Nomoto06, Sukhbold16} on the Ne/O and Mg/O panels.
The Pearson's correlation coefficients with the AtomDB results
are shown at the right-top corners on each panel.
\label{fig:tempabund}}
\end{figure*}

Here, we show trends in the metal content of various halos in our sample.
One of the most common indicators characterizing the gas property
is of course $\langle kT \rangle$;
moreover, integrated mass-to-light ratio can be another diagnostic of halos \citep[e.g.,][]{NM09}.
We computed the mass parameter by integrating the electron density in the 60\,arcsec slices
with a rough assumption that the emission is from a uniform-density sphere.
Although this is not identical to the true halo mass,
such a {\it pseudo-mass parameter} (denoted by $M_{<60''}$)
is a good surrogate for the gas mass at the central part where the halo emission is dominant.
By using parameters of the Centaurus cluster,
\citet{Fukushima22} showed that the $M_{<60''}$ calculated from the RGS
is in plausible agreement ($\sim 30$ percent) with the more accurate mass estimated
by integrating the VEM profile from CCDs.
Then, we adopt the $B$-band luminosity (denoted with $L_B$) listed in Table\,\ref{tab:observation}
to calculate the mass-to-light ratio.
In Figure\,\ref{fig:tempabund}, we plot the abundances and abundance ratios
as functions of $\langle kT \rangle$ and $M_{<60''}/L_B$.

\subsubsection{Absolute abundance}
\label{subsubsec:corrfe}

A general trend is that the absolute abundances show a positive correlation to $\langle kT \rangle$
with the Pearson's correlation coefficients $\sim 0.7$ (Figures\,\ref{fig:tempabund}(a) and (c)).
While the abundance distribution for $M_{<60''}/L_B$ has a relatively large scatter,
most of the low metallicity systems are located at the low $M_{<60''}/L_B$ regime
(Figures\,\ref{fig:tempabund}(b) and (d)).
Considering ongoing SNIa enrichments in low-$M_{<60''}/L_B$ objects,
it is unreasonable that these systems show lower Fe abundance
than the high $M_{<60''}/L_B$ systems possibly contributed by primordial gas.
Similarly, the lowest O abundances for the coolest objects (Figures\,\ref{fig:tempabund}(c) and (d))
indicates that the C regime objects (globally identical to the low-$M_{<60''}/L_B$ ones) are less enriched
with O, which is hard to interpret due to present metal supplies by mass loss.

Importantly, absolute abundances with RGS spectra are suffered from continuum estimation,
especially for subkiloelectronvolt plasmas.
The systematics in such cool plasmas have also been pointed out by many authors
\citep[e.g.,][]{Matsushita97, Matsushita00, Gastaldello21, Fukushima22}.
Even the CCD spectra are contaminated by continuum emissions from two-photon and free-bound processes
by heavy metals in cool plasmas; therefore, absolute abundance measurements
in the C or low-$M_{<60''}/L_B$ regimes are challenging with the RGS spectra that are dominated by line emissions.
We suspect that this possible bias in the C regime intrinsically contributes to
the observed abundance drops in the central coolest part of a system.

\subsubsection{Ne/O and Mg/O}
\label{subsubsec:highne}

We also plot the Ne/O and Mg/O ratios that are dominantly expelled via mass-loss winds and CCSNe,
and thus, represent nucleosynthesis in a core of supergiants.
There are clear interrelations for $\langle kT \rangle$--Ne/O
and $\langle kT \rangle$--Mg/O;
as those at the C regime are higher than the solar ratios
that are uniformly observed at the W+H regime (Figures\,\ref{fig:tempabund}(e) and (g)).
In Figures\,\ref{fig:tempabund}(f) and (h),
there are no clear correlations to $M_{<60''}/L_B$ with virtually zero Pearson's correlation coefficients
compared to relations to $\langle kT \rangle$.
However, it is an enigmatic picture that O, Ne, and Mg patterns of early-type galaxies contributed
by stellar populations are completely different from the ones in the medium in groups and clusters.
For example, the Fornax cluster and NGC~1404 are a host and its member galaxy, respectively,
that belong to the same regimes and have similar patterns.
The Virgo cluster hosts NGC~4649, M49, and M87, the former two of which exhibit
the super-solar Ne/O and Mg/O ratios.

Given that O and Ne share dominantly the same origin and enrichment process,
such schisms are hard to explain by a physical process.
Unless a picture of the dust-depleted process recovers validity
after a decrease of the conflict with flat Ne/O profiles (Figure\,\ref{fig:selectprofile}),
the abundance measurement uncertainty is instead a natural solution at present.
As both AtomDB and SPEXACT give consistent ratios,
this possible uncertainty depends on $\langle kT \rangle$.
Different from the strong O~Ly$\alpha$ line,
the Ne~Ly$\alpha$ and/or Ne~He$\alpha$ lines are not resolved completely
from the Fe~L bump even with RGS (Figure\,\ref{fig:spec});
and hence, the Ne/O ratios of the coolest objects likely suffer from some systematics.
If this is the case, unreasonably high Ne/O ratios
at the center of NGC~1404 (Figure\,\ref{fig:selectprofile}(d)),
which is the coolest object in our sample, are naturally interpreted as upward-biased values.
In addition, the Mg/O ratio, uniformly high in the C regime, is also difficult to measure
due to the low relative intensity of Mg lines in subkiloelectronvolt plasmas.

\subsubsection{O/Fe and Ne/Fe}
\label{subsubsec:ratios}

Generally, abundance ratios concerning Fe show a correlation to $\langle kT \rangle$,
and have a rather weak relation with $M_{<60''}/L_B$ (Figures\,\ref{fig:tempabund}(i), (j), (k), and (l)).
In previous studies, uniform abundance and ratios among objects in the wide mass range
are reported \citep[e.g.,][]{dePlaa07, Mernier18b}.
In Figure\,\ref{fig:tempabund}(i), the O/Fe ratio
does not correlate significantly to $\langle kT \rangle$
excluding the coolest objects, which is consistent with \citet{dePlaa17b}.
While \citet{dePlaa17b} reported that the O/Fe ratio does not vary depending on gas temperature,
including subkiloelectronvolt objects, recent works with the latest atomic codes give lower O/Fe ratios
than those measured by them \citep[e.g.,][]{Mernier22}.
The lower O/Fe of cool objects, which is possibly a biased value,
is \textit{preferable under the latest codes}.

Figure\,\ref{fig:tempabund}(k) is the first robust result of the $\langle kT \rangle$--Ne/Fe relation.
The Ne/Fe ratios have a negative $\langle kT \rangle$ correlation
with $\sim -0.5$ coefficient and give a clear delimitation
of $\langle kT \rangle$\,$= 1$\,keV
while Figure\,\ref{fig:tempabund}(l) shows that the correlation between
Ne/Fe and $M_{<60''}/L_B$ is more marginal with the Pearson's coefficient $\sim -0.3$.
We found a hint of anti-correlation of Ne/Fe to $\langle kT \rangle$
different from the $\langle kT \rangle$--O/Fe relation,
which is also suggested in Figure\,\ref{fig:tempabund}(e).
As well as the Ne/O ratios, these inverse properties of O and Ne to each other
are unreasonable because of their same origin and dispersal history.

\subsubsection{Ni/Fe}
\label{subsubsec:ni}

The Ni/Fe ratios show a moderate negative correlation
to both $\langle kT \rangle$ and $M_{<60''}/L_B$ (Figures\,\ref{fig:tempabund}(m) and (n)).
When objects fall on the $M_{<60''}/L_B \gtrsim 1\,\textrm{M}_\odot/\textrm{L}_\odot$ region,
the Ni/Fe ratios are consistent with the solar ratio.
The low-$M_{<60''}/L_B$ objects with $M_{<60''}/L_B \lesssim 0.1\,\textrm{M}_\odot/\textrm{L}_\odot$
exhibit high Ni/Fe ratios larger than 2\,solar.
While we are now prevented from assessing the Ni/Fe ratio robustly
by the absence of prominent lines of Ni in the RGS band (Figure\,\ref{fig:spec}),
a high Ni/Fe ratio is also reported with CCD for the Centaurus cluster \citep{Fukushima22}.
As indicated in many previous works, the Ni/Fe ratio is sensitive to SNIa explosion models
since a dominant fraction of Ni is synthesized at a core of exploding white dwarves
\citep[e.g.,][]{DW00, dePlaa07, Hitomi17, Mernier17}.
This will be discussed more in detail at \S\,\ref{subsec:sne}.

\subsubsection{N enrichment}
\label{subsubsec:no}

\begin{figure*}[htb!]
\plotone{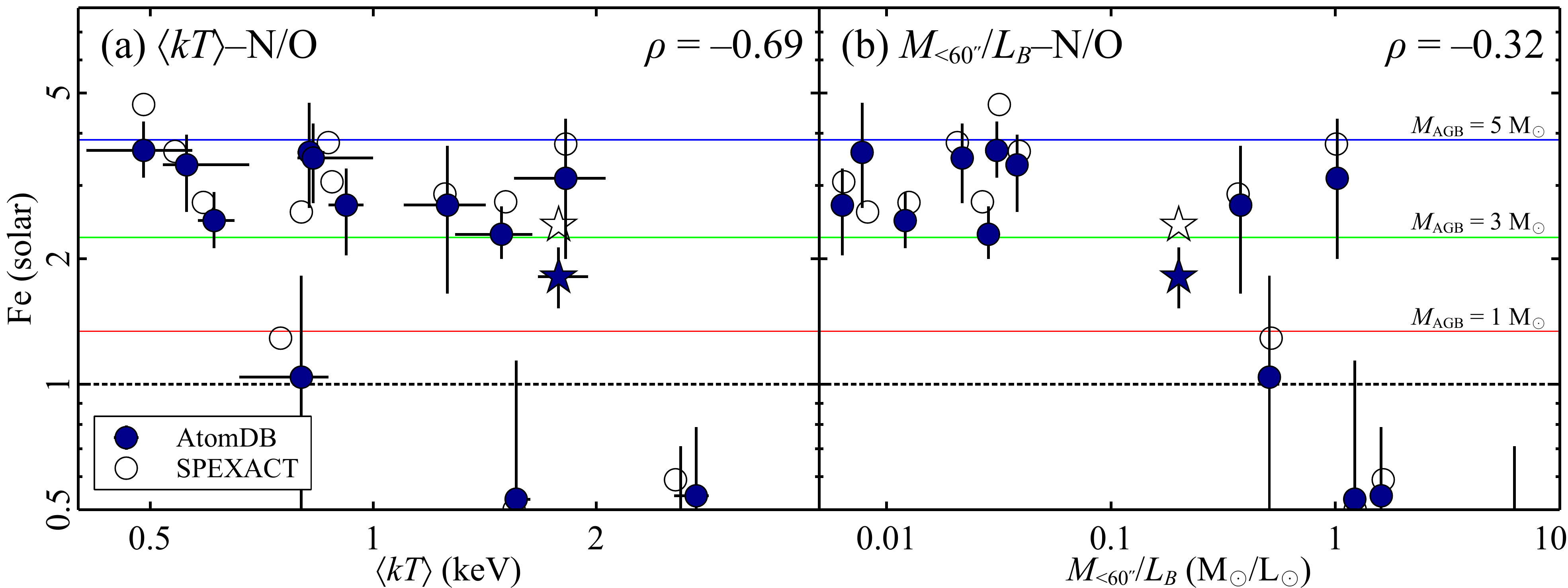}
\caption{The N/O ratios as functions of (a) $\langle kT \rangle$ and (b) $M_{<60''}/L_B$.
The filled and open circles represent the results with AtomDB and SPEXACT, respectively.
The stars indicate the results of Centaurus \citep{Fukushima22}.
Errors are shown only for the AtomDB results for viewing purposes.
The predicted ratios by \citet{Karakas10} are shown by horizontal solid lines,
assuming the initial metallicity of AGBs at each mass to the solar value.
The Pearson's correlation coefficients for the AtomDB results
are given at the right-top corners on each panel.
\label{fig:nopattern}}
\end{figure*}

The N enrichment through SNe is negligible,
and they can be instead produced and expelled by AGB stars.
We plot the N/O ratios of each object against $\langle kT \rangle$
and $M_{<60''}/L_B$ (Figures\,\ref{fig:nopattern}(a) and (b)),
comparing them to the model prediction in mass-loss winds from AGBs
($M_\textrm{initial}$\,=\,1, 3, 5$\textrm{M}_\odot$, \citealt{Karakas10}).
The N/O ratios are uniformly over-solar $\sim$\,2--4\,solar among various objects,
which are well explained by enrichments via AGBs $\lesssim 5$\,$\textrm{M}_\odot$.
Here, the dominant fraction of observed N and O is expected to be from AGB stars
that should be an ongoing enrichment channel in central regions of early-type galaxies.
Since a contribution from early CCSNe to the O enrichment is undoubtedly important
in accumulated halo history, the direct comparison of observed N/O and AGB yields is simplistic
and cannot fully rule out a possible effect of more massive AGB stars.
However, \citet{Mao19} pointed out the importance of the contribution of
low- and intermediate-mass stars ($\sim$\,0.9--7\,$\textrm{M}_\odot$) in AGB to the halo enrichment.
Additionally, \citet{Kobayashi20} also attributed the N enrichment in the Milky Way
to inter-mediate mass AGBs (4--7\,$\textrm{M}_\odot$).
Even our simple estimation gives a globally consistent result with these works.

Some hot or high-$M_{<60''}/L_B$ objects (giant clusters like Perseus) yield N/O\,$\lesssim 1$\,solar,
unlike the global trend of high N/O ratios.
One possibility is that mass-loss contribution is strong in the cool or low $M_{<60''}/L_B$ halos
compared to high N/O objects.
However, the N~Ly$\alpha$ line is weak in hot gas ($\gtrsim 2$\,keV)
and measuring the N abundance in the hot halos is harder than in the cool ones \citep{Mao19}.
Therefore, retrieving relations around the hot or high-$M_{<60''}/L_B$ regimes
from Figure\,\ref{fig:nopattern} is somewhat cautious.
A more precise trend of N/O will be validated by comparing simulation studies
with realistic assumptions of various initial mass functions (IMFs) and mass injection rates.
Spatially-resolved analysis with non-dispersive instruments will also provide a unique way to investigate
the N enrichment in early-type galaxies.

\subsection{Entire abundance patterns and SN yields}
\label{subsec:sne}

\begin{figure*}[htb!]
\plotone{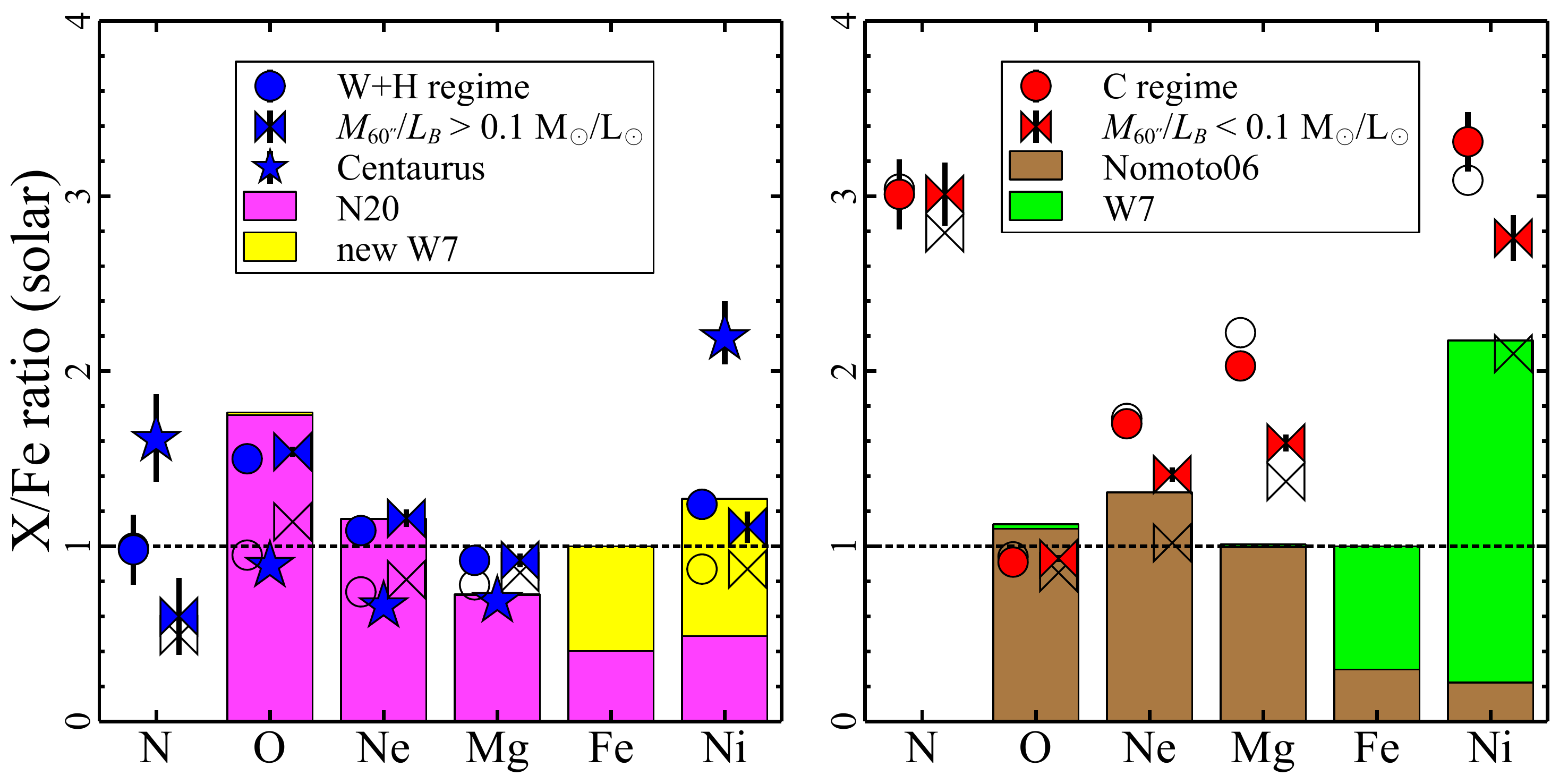}
\caption{Abundance ratio patterns obtained in central 60\,arcsec cores.
The sample is divided into two subsamples using temperature regimes (circles)
and $M_{<60''}/L_B = 0.1\,\textrm{M}_\odot/\textrm{L}_\odot$ (ties),
each of which the abundance ratios are averaged.
The stars represent the result of the Centaurus cluster given in \citet{Fukushima22}.
Filled and open plots indicate the results of AtomDB and SPEXACT, respectively.
To fit the observed pattern, we adopt two combination models of CCSN and SNIa yields:
the ``classical'' CCSN model \citep{Nomoto06} and W7 model \citep{Iwamoto99},
and N20 model \citep{Sukhbold16} and new W7 model \citep{LN18}.
Every progenitor metallicity thereof is assumed to be 1\,solar.
For viewing purpose, only two combination with the best $\chi^2$/dof
are presented for $M_{<60''}/L_B$ subsamples with AtomDB.
See Table\,\ref{tab:snfit} for the results of other regimes and combination.
\label{fig:yield}}
\end{figure*}

\begin{deluxetable*}{lrrrr}
\tablenum{4}
\tablecaption{Error-weighted average abundance ratios to Fe with AtomDB for each subsample.
Statistical variances are also given in parentheses. \label{tab:meanabund}}
\tablewidth{0pt}
\tablehead{
\colhead{} & \colhead{C} & \colhead{W+H}
& \colhead{low-$M_{<60''}/L_B$} & \colhead{high-$M_{<60''}/L_B$}}

\startdata
N/Fe & $3.0\pm 0.2$ (1.3) & $1.0\pm 0.2$ (2.7) & $3.01\pm 0.18$ (0.59) & $0.6\pm 0.2$ (2.9) \\
O/Fe & $0.91\pm 0.02$ (0.05) & $1.50\pm 0.02$ (0.10) & $0.93\pm 0.02$ (0.06) & $1.54\pm 0.03$ (0.13) \\
Ne/Fe & $1.70\pm 0.06$ (0.18) & $1.09\pm 0.04$ (0.02) & $1.41\pm 0.04$ (0.06) & $1.16\pm 0.05$ (0.05) \\
Mg/Fe & $2.03\pm 0.06$ (0.30) & $0.92\pm 0.04$ (0.06) & $1.59\pm 0.05$ (0.11) & $0.92\pm 0.04$ (0.30) \\
Ni/Fe & $3.31\pm 0.17$ (1.44) & $1.24\pm 0.09$ (0.19) & $2.76\pm 0.13$ (1.39) & $1.11\pm 0.09$ (0.15) \\ 
\enddata
\end{deluxetable*}

\begin{deluxetable*}{lrrrr}
\tablenum{5}
\tablecaption{Number ratio of SNeIa to total SNe at the best-fit combination
models for each regime using the AtomDB results. \label{tab:snfit}}
\tablewidth{0pt}
\tablehead{
\colhead{} & \colhead{C} & \colhead{W+H}
& \colhead{low-$M_{<60''}/L_B$} & \colhead{high-$M_{<60''}/L_B$}}

\startdata
\multicolumn{5}{c}{W7 \citep{Iwamoto99} + ``classical'' \citep{Nomoto06}} \\ \hline
SNIa/(SNIa+CCSN) & $0.18\pm 0.05$ & $0.18\pm 0.05$ & $0.19\pm 0.04$ & $0.16\pm 0.05$\\
$\chi^2$/dof & 11.5/3 & 15.4/3 & 5.6/3 & 20.9/3 \\ \hline
\multicolumn{5}{c}{new W7 \citep{LN18} + N20 \citep{Sukhbold16}} \\ \hline
SNIa/(SNIa+CCSN) & $0.18\pm 0.09$ & $0.11\pm 0.02$ & $0.17\pm 0.08$ & $0.11\pm 0.02$\\
$\chi^2$/dof & 32.4/3 & 1.9/3 & 25.4/3 & 2.0/3 \\
\enddata
\end{deluxetable*}

In order to summarize the entire trend of metal content for our sample,
we calculate the error-weighted average X/Fe ratios
for four subsamples of C, W+H, low-, and high-$M_{<60''}/L_B$ (Table\,\ref{tab:meanabund}).
We also show the X/Fe patterns and the best-fit linear combinations
of current SN yield models (Figure\,\ref{fig:yield} and Table\,\ref{tab:snfit}).
In the left-hand panel of Figure\,\ref{fig:yield},
the ratios are plotted for the objects belonging to the W+H and high-$M_{<60''}/L_B$ regimes.
The average ratios for the C and low-$M_{<60''}/L_B$ objects
are shown in the right-hand panel.
The result of Centaurus of \citet{Fukushima22} is not included in averages in the left-hand panel.

We reproduce these patterns with a linear combination model
of  various CCSN and SNIa nucleosynthesis calculations \citep[e.g.,][]{Simionescu19}.
Since the important ratio of SNIa to all SNe is commonly 10--20 percent with each combination,
we here give representative results with the latest calculations of
N20 model from \citet{Sukhbold16} $+$ new W7 model of \citet{LN18}.
The results with well-established ``classical'' yields of
\citet{Nomoto06} $+$ W7 model of \citet{Iwamoto99} are also reported for comparison.
The IMF of \citet{Salpeter55} is used to average over
various progenitor masses for CCSNe.
We exclude yield calculations for SNeIa with a sub-Chandrasekhar mass white dwarf
because specific elements to them like Ca or Cr are completely absent from patterns in Figure\,\ref{fig:yield}.

The N/Fe ratios are also plotted in Figure\,\ref{fig:yield} for referential results.
One can include AGB contribution in the combination model
to reproduce the N abundance. However, the N enrichment by SN explosions is negligible,
and other elements used in Figure\,\ref{fig:yield} are little affected
by this additional component \citep[][]{Mao19}.
Moreover, the SN yield fitting method abandons information about timing evolution,
making it difficult to separate each SN contribution from the AGB yield
as the ejecta of early CCSN and SNIa had formed stars that are now in the AGB.
Therefore, the N/Fe ratios are not used to fit the patterns in this study.

\subsubsection{Hot or High-$M_{<60''}/L_B$ objects}
\label{subsubsec:hothigh}

In the left-hand panel of Figure\,\ref{fig:yield},
both patterns of the W+H and high-$M_{<60''}/L_B$ regimes are in plausible agreement
for O/Fe, Ne/Fe, Mg/Fe, and Ni/Fe.
The O/Fe, Ne/Fe, and Mg/Fe ratios are close to the solar ratios.
This pattern is reproduced well by the latest N20 model of \citet{Sukhbold16},
also suggested in the Ne/O and Mg/O plots (Figures\,\ref{fig:tempabund}(e) and (g)).
We found that the Ni/Fe ratio is also about 1\,solar, especially for SPEXACT,
and that it prefers the new W7 calculation by \citet{LN18}.
The sub-solar N/Fe ratio of high-$M_{<60''}/L_B$ objects compared to the W+H ones,
still cautious to directly accept (see \S\ref{subsubsec:no}),
implies the SNIa effect in systems with long-timescale enrichment.
Such a trend is also predicted by simulation study \citep[e.g.,][]{Kobayashi20}.
For these patterns, the number ratio of SNeIa to total SNe is $\sim 11$ percent (Table\,\ref{tab:snfit}),
which contributes to 60 percent of observed Fe.
These results are globally consistent with the results for the Perseus cluster \citep{Simionescu19}.
While the two codes give consistent results within atomic code uncertainties
(see \S\ref{subsec:atomcode}), the O/Fe ratio with SPEXACT is closer to the solar value
than AtomDB, which is also indicated in \citet{Simionescu19}.

Notably, the O/Fe, Ne/Fe, and Mg/Fe ratios of Centaurus are
slightly smaller than the other objects in the same regime.
\citet{Fukushima22} suggests a significant contribution of SNIa in this object
compared to other cool-core systems.
High N/Fe and Ni/Fe ratios also distinguish the Centaurus cluster from the others.
This object likely shows an intermediate property of enrichment channels
between the two temperature regimes (e.g., mass-loss winds, SNe, primordial gas),
also implied by medium $\langle kT \rangle$ and $M_{<60''}/L_B$
in our sample (Figure\,\ref{fig:tempabund}).
Interestingly, a recent cosmological simulation predicts sub-solar ratios
for O/Fe, Ne/Fe, and Mg/Fe \citep{Fukushimak22},
which is consistent with the pattern of Centaurus.
While there can still be some uncertain factors in the simulation
(e.g., the delay time distribution of SNIa, mass-loss rate, IMF),
traditional linear combination modeling is too simplistic for discussing halo enrichment.
Closer cooperative studies of observing, simulating, and theoretical modeling
are more important in the next era of high-resolution spectroscopy.

\subsubsection{Cool or Low-$M_{<60''}/L_B$ objects}
\label{subsubsec:coollow}

It is easy to find that the C or low-$M_{<60''}/L_B$ regime shows the solar O/Fe ratio
with both AtomDB and SPEXACT (right-hand panel of Figure\,\ref{fig:yield}),
which makes these objects prefer the ``classical'' CCSN model of \citet{Nomoto06}.
On the other hand, the Ne/Fe and Mg/Fe ratios are globally higher than the O/Fe, especially with AtomDB.
This is also implied in the plots of Ne/O and Mg/O (Figure\,\ref{fig:tempabund}(e), (f), (g), and (h)).
High Ne/Fe and Mg/Fe ratios are not reproduced well by our SN combination models
with high $\chi^2$/dof (Table\,\ref{tab:snfit}).
In particular, the combination of the latest SN models \citep{Sukhbold16, LN18}
for the C regime yield the worst $\chi^2$/dof value among all trials (Table\,\ref{tab:snfit}).
The Ni/Fe ratios are higher than 2\,solar, sometimes exceeding the high predictive value
of the W7 model by \citet{Iwamoto99}.
Super-solar N/Fe ratios are well explained by mass-loss winds from 3--4$\textrm{M}_\odot$ stars in AGB,
consistent with the picture provided in \S\,\ref{subsubsec:no}.
The best-fit combination model suggests that SNIa/SN is $\sim 20$ percent
(Table\,\ref{tab:snfit}) and SNeIa produce 70 percent of Fe.

\subsection{Future prospects}
\label{subsec:future}

\begin{figure}[htb!]
\plotone{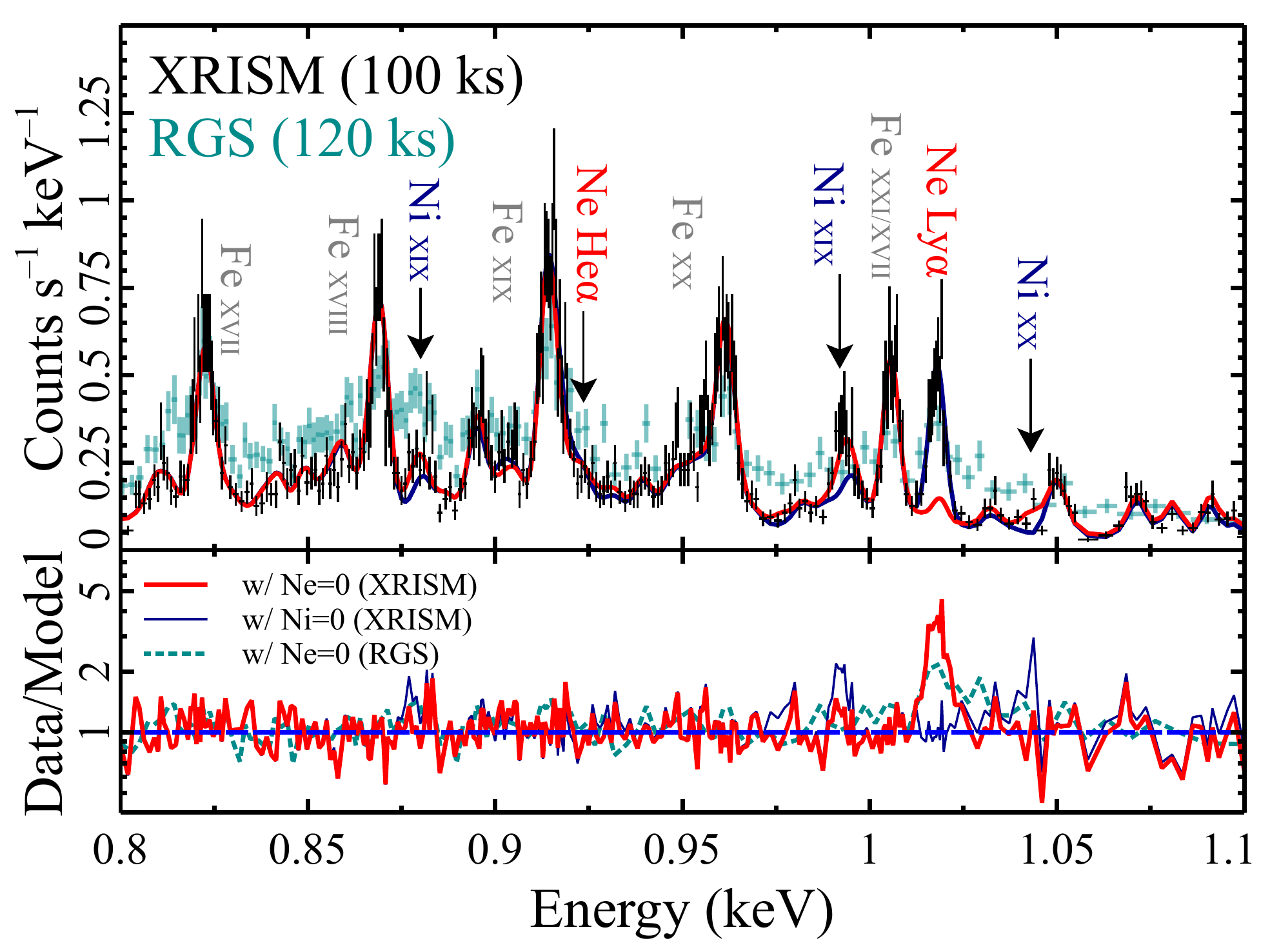}
\caption{XRISM simulated spectrum of the central arcmin scale of NGC~4649.
The current RGS data within the $60''$ region is also plotted in dark cyan.
The red and dark blue lines represent the best-fit models without Ne and Ni lines, respectively.
Prominent emission lines are marked on the upper panel.
\label{fig:simspec}}
\end{figure}

As discussed in \S\,\ref{subsec:atomcode} and \ref{subsec:sne},
current atomic codes and emission models are still limited and endeavored.
Hence, future X-ray missions with high-resolution spectroscopy
like XRISM \citep{XRISM20} and Athena \citep{Nandra13}
are strengthening more and more importance.
For example, the Resolve instrument on XRISM obtained at a great $< 7$\,eV spectral resolution
will offer fully resolved spectra within the Fe-L complex band even in subkiloelectronvolt plasmas.

In Figure\,\ref{fig:simspec}, we show the simulated Resolve spectrum of NGC~4649
that exhibits $\langle kT \rangle$\,$\sim 0.8$\,keV and the strongest Ne~Ly$\alpha$ emission
among the coolest three objects in Table\,\ref{tab:parameter}.
In the simulated spectrum, the Ne~Ly$\alpha$ is completely resolved
from a close Fe-L line with great sharpness and significance.
Determining the Ne abundance with great accuracy more confidently addresses
our proposed Ne issues with the current RGS data
(\S\,\ref{subsec:flat}, \ref{subsubsec:highne}, and \ref{subsubsec:ratios});
for example, whether the high Ne/O ratio in subkiloelectronvolt halos is real or an artifact.
This is important for determinately preferring or rejecting the dust formation scenario in early-type galaxies.
We also expect that observing cool and metal-rich objects with O, Ne, and Mg lines
impacts the theoretical research of CCSNe and the nucleosynthesis in supergiants.

Furthermore, we can unearth some Ni-L emissions that are hidden under the continuum
with the RGS spectrum and detect them as relatively sharp line structures.
The clear detection of Ni-L lines in cool plasma allows us to measure the Ni abundance robustly,
even though the Ni K-shell emissions are absent in observed objects.
As indicated in \S\,\ref{subsubsec:ni}, the Ni/Fe ratio is super-solar in such cool plasma systems.
Measuring the Ni/Fe ratio with clear emission lines in these cool objects will significantly improve
our current unclear knowledge of SNIa in early-type galaxies
that are less enriched and assembled than are groups or clusters.
Studying SNeIa in early-type galaxies is a good complement to the Hitomi results
of the Perseus cluster \citep{Hitomi17, Simionescu19} that is one of the most massive objects.
The anomalous Ni/Fe ratio in the Centaurus cluster will also be discussed in more detail.

Finally, worth stressing is that spectra with well-resolved Fe- and Ni-L lines are ideal \textit{laboratories}
to test the atomic codes like AtomDB and SPEXACT.
While great efforts in laboratory measurements have improved the codes
and reduced their uncertainties \citep[e.g.,][]{Gu19, Gu20, Gu22a},
some disagreements between the two codes are controversial in current observational studies,
especially around Fe-L lines in cool plasma \citep[e.g.,][]{Gastaldello21, Fukushima22}.
In tandem with this progressive laboratory physics, high-resolution spectroscopy
will help us to \textit{iron} the persistent wrinkles out of the atomic codes.

\section{Conclusions}
\label{sec:conclusions}

We have analyzed the RGS data of 14 nearby early-type galaxies including BCGs in groups and clusters,
and measured the radial metal distribution, abundance pattern, and gas mass in these objects.
The results are summarized as follows:

\begin{itemize}
\item[--] Radial profiles of the O/Fe, Ne/Fe, and Mg/Fe ratios are generally flat
at the center of each object, irrespective of the presence of Fe abundance drops.
This result is hard to explain by the dust depletion scenario
although the only exception of NGC~1404 shows centrally enhanced Ne and Mg.

\item[--] We find systematic differences of atomic codes up to 10, 30, and 40 percent
for determining $\langle kT \rangle$, absolute abundances, and relative abundance ratios,
which is consistent with the previous report \citep{Mernier20a}.
These differences are not depending on the temperature regimes.

\item[--] The abundances and abundance ratios correlate more strongly
to $\langle kT \rangle$ than to $M_{<60''}/L_B$.
The Ne/O and Mg/O ratios are higher than the solar ratio
for the objects of $\langle kT \rangle$\,$<$\,1\,keV.
We suspect that high abundance ratios for cooler objects are systematically biased upward.
The Ni/Fe ratio is also high for the cool objects with AtomDB but globally consistent with solar with SPEXACT.

\item[--] The N/O ratios are uniformly high $\sim 2$--$4$\,solar.
Assuming that observed N and O are dominantly from ongoing enrichment,
this is well reproduced by mass-loss winds from $3$--$5\textrm{M}_\odot$ stars in the AGB.
This is consistent with the N enrichment expected in Milky Way \citep{Kobayashi20}.

\item[--] Abundance patterns of the hot or high-$M_{<60''}/L_B$ systems are consistent with the solar composition.
For the cool and low-$M_{<60''}/L_B$ objects, the super-solar Ni/Fe ratios are likely overestimated.
The high Ne/Fe and Mg/Fe ratios are observed only for the pattern of cool objects,
which also supports the presence of measurement uncertainties in these abundances.

\item[--] A next leap will be achieved by non-dispersive and high-resolution
spectroscopic missions like XRISM and Athena,
which constitute unique ways of measuring metal abundances in a subkiloelectronvolt plasma.
The robust determination of the Ne and Ni abundances at unprecedented accuracy
provides us invaluable information for SN contribution to the enrichment of early-type galaxies.

\end{itemize}


The authors sincerely thank the anonymous referee for
the constructive comments and suggestions,
which have dramatically improved our manuscript.
K.F. acknowledges financial support from the Grants-in-Aid for Scientific Research (KAKENHI) program
of the Japan Society for the Promotion of Science (JSPS)
grant Nos.~21J21541 and 22KJ2797 (Grant-in-Aid for JSPS Fellows).


{
\facility{XMM}
\software{\textsc{xspec} \citep{Arnaud96}, \textsc{spex} \citep{Kaastra96},
\textsc{spex\_to\_xspec} (\url{https://github.com/jeremysanders/spex_to_xspec}),
\textsc{python} (\url{https://www.python.org})
}


\bibliography{paper_ref}{}
\bibliographystyle{aasjournal}



\end{document}